\RequirePackage{fix-cm}
\documentclass[smallextended]{svjour3}       
\smartqed  
\usepackage{graphicx}
\usepackage{mathptmx}  
\usepackage{color}
\usepackage{url}
\usepackage[T1]{fontenc}

\usepackage{mathptmx}
\usepackage{url}
\usepackage{multirow} 
\usepackage[table,xcdraw]{xcolor}

\usepackage{cancel}

\usepackage[colorlinks=true, linkcolor=blue, citecolor=blue, urlcolor=blue]{hyperref}

\begin{document}

\title{On the Need to Monitor Continuous Integration Practices 
}
\subtitle{An Empirical Study}


\author{Authors}
\author{Jadson Santos         \and
        Daniel Alencar da Costa \and
        Shane McIntosh    \and
        Uir\'a Kulesza 
}


\institute{
  J. Santos \at
  Federal University of Rio Grande do Norte \\
  Natal, Brazil\\
  \email{jadson.santos@ufrn.br} \and 
  D. A. da Costa \at
  University of Otago \\
  Dunedin, New Zealand \\
  \email{danielcalencar@otago.ac.nz} \and 
  Shane McIntosh \at
  University of Waterloo \\
  Waterloo, Canada \\
  \email{shane.mcintosh@uwaterloo.ca} \and 
  U. Kulesza \at
  Federal University of Rio Grande do Norte \\
  Natal, Brazil \\
  \email{uira.kulesza@ufrn.br}
}

\date{Received: date / Accepted: date}

\maketitle

\begin{abstract}
One of the crucial activities in software development is monitoring. It plays a vital role in verifying the proper implementation of processes, the identification of errors, and the discovery of opportunities for improvement. Continuous Integration (CI) encompasses a set of widely adopted practices that enhance software development. However, there are indications that developers may not adequately monitor CI practices. Hence, this paper explores developers' perceptions regarding monitoring CI practices. To achieve this, we first defined metrics associated with CI practices. Then, we conducted a series of analyses to evaluate how these metrics are perceived in projects or by their developers. As a starting point, we perform a Document Analysis to assess developers' expressed need for practice monitoring in pull request comments generated by developers during the development process. After that, we conduct a survey of 28 developers from 121 open-source projects to understand the perception of the significance of monitoring seven CI practices in their projects. Finally, we triangulate the emergent themes from our survey by performing a second Document Analysis to understand the extent of monitoring features supported by existing CI services. Our key findings indicate that: 1) the most frequently mentioned CI practice during the development process is ``Write automated developer tests'', which is associated with the metric ``Coverage'' (> 80\%), while ``Don't commit broken code'' and ``Fix Broken Builds Immediately' present notable opportunities for monitoring CI practices; 2) developers do not adequately monitor all CI practices and express interest in monitoring additional practices; and 3) the most popular CI services currently offer limited native support for monitoring CI practices, requiring the use of third-party tools. Our results lead us to conclude that monitoring CI practices is often overlooked by both CI services and developers. Using third-party tools in conjunction with CI services is challenging, they monitor some redundant practices, and they still fall short of fully supporting CI practices monitoring. Therefore, CI services should implement CI practice monitoring, which would facilitate and encourage developers to monitor them.

\end{abstract}

\section{Introduction} 

Continuous Integration (CI) is a set of software development practices where change sets that teams produce are built and tested as they are submitted for inclusion in a repository of records. In many settings, CI is invoked several times per day. Each invocation triggers an automated build, which can ideally detect errors as early as possible \cite{Duvall2007}. Over the past two decades, there has been a growing acknowledgement of the benefits derived from increasing the frequency of integration and release activities. The term ``Continuous
$^\star$'' (continuous star) was introduced by Fitzgerald et al.~\cite{fitzgerald2014continuous} to describe this collection of continuous activities.

Although CI has become a popular topic in research and industry \cite{StakeholderPerceptions2015,ReplicationCIPainPoints2019,TradeOffsCI2017,ImpactCICodeReviews2020,hilton2016usage,benefitsCIIndustry2013,elazhary2021uncovering,VasilescuQualityProdOutcomes2015,JoaoHelis2018,santos2022investigating,ghaleb2019empirical,coverageImpac2017}, it is not always the case that development teams fully embrace the spirit of CI and, as such, tend not to enjoy its full benefits. Development teams may perform this so-called ``{\em CI Theater}''\cite{CITheater2019} where they have the illusion of practicing CI without fully adopting all its practices, often due to the costs and challenges associated with implementing CI. This lack of complete adoption of CI practices may lead to unhealthy practices, as discussed by Felidré et al. \cite{CITheater2019}. Indeed, Hilton et al.~\cite{hilton2016usage} and Elazhary et al.~\cite{elazhary2021uncovering} asked developers why they did not (fully) adopt CI, observing recurring themes, such as the trade-off between test effectiveness and build speed~\cite{ghaleb2019empirical}. 

Despite the central role that CI plays in software development, it remains unclear to what extent stakeholders are aware of the maturity {\em} of their CI workflows. We believe that such an awareness would require active {\em monitoring of CI practices}. For example, a company studied by Elazhary et al.~\cite{elazhary2021uncovering} admittedly runs tests inefficiently due to project constraints on the available test infrastructure. In turn, such choices prolong build durations, which can hinder organizational agility \cite{chen2020buildfast}. However, what if the monitoring of build durations could help the said company make the business case to invest in additional test infrastructure? Without awareness of the current CI workflow, such investment decisions tend to be made intuitively and, as such, are prone to limitations and biases. Indeed, the recent debut of \textsc{Metrics GitHub Actions} service \footnote{https://github.blog/2023-07-19-metrics-for-issues-pull-requests-and-discussions/} suggests that monitoring metrics related to CI is a feature that CI providers believe is practical and relevant.

As such, our research focuses on a set of questions regarding the monitoring of CI practices, such as: do development teams monitor their CI practices? Do they perceive the monitoring of CI practices as important? Which monitoring capabilities do state-of-the-art CI services provide? {\em Continuous monitoring} facilitates the timely identification of quality-of-service issues, including performance degradation, and the fulfillment of Service Level Agreements (SLAs)~\cite{fitzgerald2014continuous}. Typically, monitoring focuses on application performance (e.g, CPU, memory, network, and disk) and infrastructure
(e.g, physical servers, virtual machines, databases, network infrastructure), nonetheless, our work focuses on monitoring the software integration process, which is often overlooked.

To better understand the importance and prevalence of monitoring in the context of CI practices, we conduct a series of qualitative analyses. We start by investigating whether developers express the need for monitoring CI practices in their project repositories. To do so, in RQ1, we perform a Document Analysis~\cite{bowen2009documentanalysis} on 4,892 Pull-Request (PR) comments to identify the situations where developers have expressed the need for monitoring CI practices. 

Next, given the insights obtained, we triangulate the emergent themes from RQ1 using a questionnaire that we issue to 28 developers. Finally, we perform another Document Analysis of the official documentation of current state-of-the art CI services to investigate the currently available monitoring capabilities of such CI services. We structure our observations by addressing the following Research Questions (RQs):

\begin{itemize}

\item \textbf{RQ1: Do developers express the need to monitor CI practices in PRs?} Our findings show that the main needs expressed by developers are related to monitoring ``Don't commit broken code'' and ``Fix Broken Builds Immediately'' since they often encounter build failures in PR comments. We also discover that ``Write Automated Developer Tests'' is the most monitored practice (83\% of the monitoring opportunities found in PR comments, suggested the monitoring of Coverage metric), and the \textsc{CodeCov} tool is the most commonly used tool for monitoring this practice (86\% of coverage monitoring in PR comments used this tool).

\item \textbf{RQ2: What is the perceived importance of monitoring CI practices?} The majority of our respondents $\frac{20}{28}$ ( or 71\%) consider the CI practices that we summarize in the questionnaire to be important answering ``Yes'' to the open-ended question: \textit{``Do you think it is important for your project to measure the evolution of these practices over time? Why?''}, and would like to monitor them if given the opportunity. However, since CI services do not provide built-in support, most of our respondents $\frac{19}{26}$ ( or 73\%) report that they monitor coverage using the \textsc{CodeCov} tool and build metadata (e.g. build duration and build status) using GitHub Actions.


\item \textbf{RQ3: Which monitoring features are currently supported by existing CI tools?} CI services currently offer limited built-in support for monitoring CI. Only the Gitlab documentation suggests an implementation of a dashboard for monitoring metrics related to continuous activities. Build metadata tracking are the most commonly supported monitoring metrics. There are various complementary tools that can enhance this monitoring. However, they often come with additional costs and complexity, with a predominant focus on fundamental information about the performance of the CI pipeline.

\end{itemize}

Our findings highlight that ``Coverage'' (that is associated with ``Write Automated Developer Tests'' practice) was the most frequently monitored metric $\frac{258}{306}$ ( or 80\%), with \textsc{CodeCov} being the main tool used in most projects $\frac{206}{258}$  ( or 79\%), a result reinforced by the participants of our survey. The primary opportunities lie in monitoring ``Don't commit broken code'' and ``Fix Broken Builds Immediately''. A significant majority of participants in our survey $\frac{20}{28}$ ( or 71\%) demonstrated interest in monitoring CI practices. Furthermore, we observed that CI services do not provide built-in support to monitor CI practices and the adoption of third-party tools for monitoring CI practices still presents challenges that need to be overcome.

\section{Assessment of Continuous Integration} \label{Related}

In this section, we situate our study with respect to previous works that assessed CI and its impacts.

\subsection{CI Adoption and Its Impacts} \label{practices}

There is a considerable body of literature on the topic of CI adoption, exploring its benefits and costs. However, most of these studies have concentrated on evaluating CI adoption in its entirety (i.e., evaluating effects prior to and after adopting  CI). Many of these studies lack clear criteria for defining the use of CI ~\cite{EliezioRevisaoSistematica2022}. Among the studies lacking precise criteria, the most commonly applied criterion, based on interviews, surveys, or other data sources (e.g., build logs), was the adoption of a CI service.

Hilton et al.~\cite{hilton2016usage} conducted an analysis of 34,544 open-source projects from GitHub and surveyed 442 developers. These developers responded to 14 key questions regarding the usage, costs, and benefits of CI. Notable findings from Hilton et al. include: (i) projects using CI release more than twice as often as those not using CI. (ii) the average build time is just under 500 seconds, and (iii) they observed that successful builds run faster than failed builds. Vasilescu et al. \cite{VasilescuQualityProdOutcomes2015} developed a study to discern the effects of CI adoption in quality and productivity outcomes. They collected a dataset of 246 GitHub projects which at some point in their history added the Travis CI to the development process. They found that after adopting the Travis CI, teams are significantly more effective at merging pull requests submitted by core members. They also report that core developers in teams using CI are able to discover significantly more bugs than in teams not using CI. Lastly, Zhao et al. \cite{ZhaoImpactCI2017} reported on a study involving a large sample of GitHub open-source projects that had adopted Travis CI, how development practices changed after transition to using Travis CI. The main results of their study are: (i) the increasing number of merge commits, (ii) the ``commit small'' guideline is not followed by all projects, (iii) an increasing trend in the number of closed pull requests, (iv) the pull request latency increases despite the code changes becoming smaller, and (v) test suite sizes seem to increase after adopting Travis CI. 

Our study differs from previous ones in two main aspects. Firstly, we exclusively analyzed projects that had already adopted CI. Secondly, our work focuses on evaluating the impact of monitoring specific CI practices on projects. Through a qualitative study, we surveyed various aspects of CI practices, gathering insights from the perspective of project developers.

\subsection{CI Practices and Its Impacts} \label{practices}

Only recently have some studies started to conduct more in-depth evaluations of CI adoption by measuring the  impact of specific CI practices~\cite{fowler2006continuous}. CI practices consist of technological choices that stakeholders must make and habits that developers must form to improve collaboration and code quality. These practices aim to integrate code changes from multiple developers into a shared repository frequently and build and test the software automatically to identify potential integration issues. By adopting CI practices, development teams can catch integration issues early, identify and resolve conflicts, and ensure that the software remains functional and stable throughout the development process. This leads to improved collaboration, faster delivery of features, and higher code quality \cite{hilton2016usage,betz2013implementing}.  

Duvall et al. \cite{Duvall2007}  recommended seven best practices for individuals and teams running CI on a project: i) Commit Code Frequently, ii) Do not Commit Broken Code, iii) Fix Broken Builds Immediately, iv) Write Automated Developer Tests, v) All Tests and Inspections must Pass, vi) Run Private Builds, and vii) Avoid Getting Broken Code. While recommendations are plentiful, there is still no consensus on the set of CI practices that a project should follow, and what specific values teams should strive for (e.g., number of commits per day or hour). Indeed, the ideal extent to which these practices are employed may vary from project to project, depending on their domain, the quantity of features being produced/maintained, and/or the size of its user base. This variation in CI adoption makes monitoring even more challenging. In this paper, we analyze the following measures that align with Duvall et al.'s \cite{Duvall2007} seven best practices for CI: (i) Commit Per Weekday, (ii) Coverage, (iii) Build Duration, (iv) Build Activity/Build-frequency, (v) Build Health, (vi) Time to Fix a Broken Build, and (vii) Comments Per Change/Comments Per PR.

 
Santos et al. \cite{santos2022investigating} collected data on CI practices from 90 relevant open-source projects over a 2-year period and performed a quantitative analysis to investigate whether these practices are related to higher project productivity or quality. Furthermore, Elaszhary et al. \cite{elazhary2021uncovering} investigated the extent to which three software organizations implement ten Continuous Integration (CI) practices defined by Fowler \cite{fowler2006continuous}. They explored the benefits these practices bring and the challenges encountered during their implementation. This inquiry was conducted through a multiple-case study with mixed methods, focusing on three small to medium-sized software-as-a-service organizations. Their study reveal which of the ten CI practices the three companies adopted, how and why they adopted them, the benefits they perceived, and the challenges they encountered. Moreover, their findings emphasizes the importance of the context in which CI is applied. Many aspects of CI vary depending on how it is implemented. While organizations broadly adopt CI practices, they adapt them based on perceived benefits, project context, and the CI tools they use. The study maps out the benefits and challenges each company faces when adopting CI practices, focusing on areas where companies differ significantly and where these differences reveal important considerations. Among the key findings, the paper highlights: (i) addressing merge conflicts has a higher priority than the unclear benefits of using a single source repository, (ii) the perceived effort required for integration testing outweighed its perceived benefits, (iii) data-centric projects tend to have longer builds, as comprehensive testing takes priority over shorter build times, and (iv) for some organizations, enforcing deployment privileges to enhance security was prioritized over fostering a sense of ownership in the deployment process. The authors emphasize the need for researchers to consider these variations when studying projects that claim to use CI. Their work also underscores the importance of measuring CI practices to better understand the context in which they are adopted. 

Our work distinguishes itself from the study conducted by Santos et al. \cite{santos2022investigating}  by evaluating the impact of these practices on developers from a broader set of perspectives, rather than solely focusing on their effects on project productivity or quality. Additionally, we are conducting a more comprehensive study, examining the need for monitoring CI practices on a larger number of projects compared to the study by Elaszhary et al. \cite{elazhary2021uncovering}, through a survey applied to 121 open-source projects. These works come closest to ours, but our work is unique because no one has previously checked the need to monitor CI practices.

\subsection{DevOps practices monitoring} \label{DevOps-practices}

DevOps \cite{ebert2016devops} is an approach that emphasizes collaboration, communication, and integration between development (Dev) and operations (Ops) teams to enable faster and more efficient software delivery. Continuous Integration (CI), is one of the primary stages of DevOps \cite{yarlagadda2018understanding}. Concerning DevOps, CI serves as a foundational step that allows teams to achieve continuous delivery and continuous deployment \cite{karamitsos2020applying,lwakatare2016relationship}. It sets the stage for other DevOps practices such as automated testing, infrastructure as code, continuous deployment, continuous delivery and continuous monitoring \cite{vadapalli2018devops}. Together, CI and other DevOps practices promote collaboration, reliability, stability, resilience, and efficiency in software development and operations \cite{yarlagadda2018understanding}.

To monitor adherence to DevOps best practices, a set of measures have been proposed. DevOps Research and Assessment (DORA) strive to measure software development and delivery performance \cite{DoraMetricsLeanIX2023}. The four DORA metrics are: (i) Deployment Frequency (DF), (ii) Lead Time for Changes (LT), (iii) Mean Time to Recovery (MTTR) and (iv) Change Failure Rate (CFR). These metrics are designed to help organizations to monitor their performance with respect to DevOps best practices, identify areas for improvement, and track progress over time. They provide valuable insights into the speed, stability, and quality of software delivery, enabling teams to make data-driven decisions and continuously improve their DevOps performance.

Sallin et al. \cite{sallin2021measuring} conducted a comprehensive literature review to evaluate the feasibility of automatically measuring the four DORA Metrics. They found no scientific literature that specifically investigated the automatic measurement of these metrics. However, their review of gray literature uncovered sixteen articles discussing aspects of automatic measurement. Additionally, they conducted a case study with a Scrum team of 10 people to assess the value of automating the measurement of DORA metrics. 

While evaluating the current practices supported by CI tools in our work (RQ3), we identified several metrics that extend beyond the traditional scope of CI, including the DORA metrics. In complement to the work of Sallin et al. \cite{sallin2021measuring}, we assessed the support for these metrics in leading CI services and third-party monitoring tools.

\subsection{Continuous Integration Maturity Model}

When it comes to adopting CI practices, the level of adherence to these practices can also vary over time and across projects. For example, projects that use CI tools but do not commit frequently while maintaining fast builds, are not in full adherence with CI best practices. Such cases have been referred to as ``CI Theater''~\cite{CITheater2019}. The gray literature has defined a CI maturity model \cite{DeliveryMaturityModel2013} in five levels, which has: (i) base, (ii) beginner, (iii) intermediate, (iv) advanced and (v) expert levels. The model also defines five categories that represent key aspects to consider when adopting CI: (i) Culture \& Organization, (ii) Design \& Architecture, (iii) Build \& Deploy, (iv) Test \& Verification, and (v) Information \& Reporting.  This definition from the gray literature proved to be more practical and easier to understand when adopting CI for a project. We incorporated these definitions from gray literature into our survey, to capture developers' perceptions of the CI maturity level in their projects. 

Santos Jr et al.~\cite{MonalessaAmIGoingtoHaven2021} propose a diagnostic instrument designed to facilitate the identification of the extent of Continuous Software Engineering (CSE) adoption. Their diagnostic approach is rooted in developers' perceptions of the practices implemented within their respective companies. This diagnosis is executed through the utilization of the Stairway to Heaven model (StH), which comprises five distinct stages: (i) Traditional Development, (ii) Agile Organization, (iii) Continuous Integration, (iv) Continuous Deployment, and (v) R\&D as an Innovation System. The Diagnosis Questionnaire consists of practices identified from both existing literature and the authors' practical experience. These practices are systematically categorized and aligned with the various stages. Users are required to indicate the level of CSE practice adoption within their organizations.

While their study relies on developers' questionnaire responses regarding practices within their companies, our approach is more objective. We collected data and calculated the values of CI practices, subsequently verifying project developers' perceptions based on these calculated CI practice values.

The objective of our work is to comprehensively understand the necessity of monitoring CI practices from three perspectives: (i) investigating whether developers express a need for monitoring CI practices within their project repositories, (ii) gathering insights from project developers on the need to monitor CI practices, and (iii) examining the currently available monitoring capabilities of CI services to provide a holistic view of the state of CI monitoring in the industry.

\section{Methodology} \label{Methodology} 

In this section, we present our research questions. For each question, we explain our motivation and methodological approach.

\subsection*{\textbf{RQ1: Do developers express the need to monitor CI practices in PRs?}}  \label{sub:rq1}

\textbf{\textit{\underline{Motivation:}}} The starting point of our research is to understand whether developers express their concern regarding monitoring the adherence level to the CI practices in their projects. Understanding developers' concerns can help the developers to identify areas for improvement and ensure that best practices are being followed. Moreover, this can lead to further research initiatives aimed at developing better tools, methodologies, and best practices for CI monitoring in software development projects. To identify CI practices, we examine the CI metrics associated with them in developer discussions reflected in pull request (PR) comments. We have two main goals with this analysis:

\begin{itemize}
\item  Identify opportunities/needs for monitoring certain CI practices
\item  Identify evidence of what CI practices the projects are currently monitoring
\end{itemize}

\textbf{We collected data between August and October 2022.}

\begin{figure}[!htbp]
  \centering
  \includegraphics[width=\linewidth]{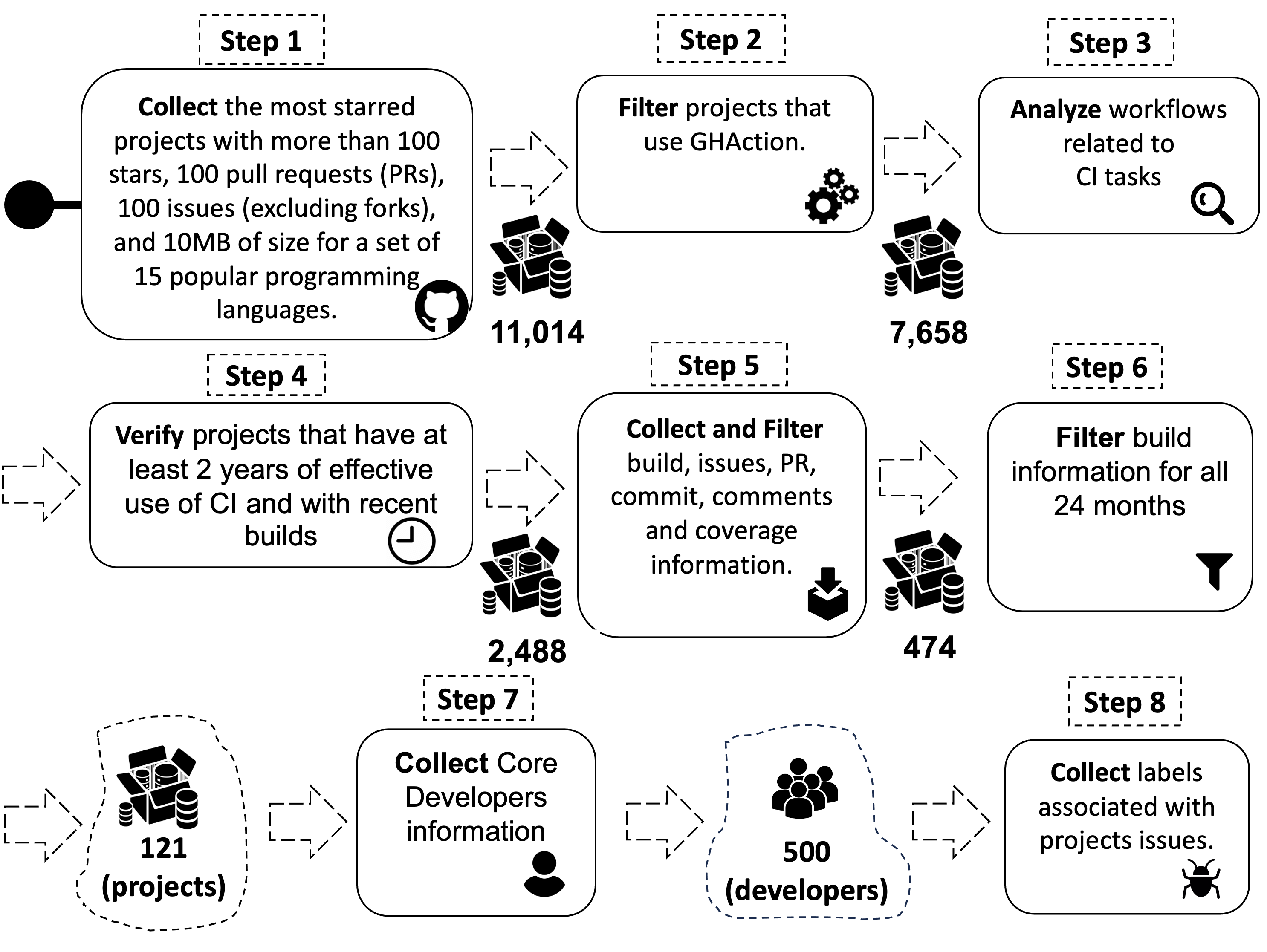}
  \caption{Selection Data Overview}
  \label{fig:approach_overview}
\end{figure}

\hfill \break
\textbf{\textit{\underline{Approach:}}} To collect a representative sample of PR comments that provide insight into how developers discuss CI monitoring, we first need to select a representative sample of non-trivial open-source projects. Following an approach similar to other studies \cite{VasilescuQualityProdOutcomes2015,Vassallo01,JoaoHelis2018,santos2022investigating},  we outlined the selection process steps in Figure~\ref{fig:approach_overview}. In the following paragraphs, we will detail the steps taken to select the projects for our study.

In \textbf{Step 1} of our study, we search for the most starred repositories with more than 100 stars on \textsc{GitHub} across a set of 15 popular programming languages: Java, JavaScript, C\#, Python, PHP, TypeScript, C, Go, C++, Kotlin, Ruby, Rust, Swift, Scala, and Objective-C. This can be achieved using the GitHub Search REST API\footnote{https://docs.github.com/en/rest/search/search}, which allows retrieving projects in a specific language, ordered by the number of stars. For this study, we used the Snooper project\footnote{https://github.com/jadsonjs/snooper}, which provides several pre-imple-mented queries and abstracts calls to the GitHub REST API. We exclude repositories that were forks and that were smaller than 10MB. Our search yielded a pool of 41,469 repositories. We filter out repositories with no more than 100 pull requests (PRs) and 100 issues, resulting in a reduced set of 11,014 projects. We applied these filters to ensure the selection of relevant projects with sufficient activity and data for our study, thus we focused on projects that provided enough information to analyze CI practices meaningfully. The threshold value of 100 was chosen based on previous studies \cite{VasilescuQualityProdOutcomes2015,JoaoHelis2018,santos2022investigating}. Our threshold of 10MB for repository size was also based on prior research \cite{MarcosOliverira2017,DEOLIVEIRA2019110420,ConvolutionalAttentionNetwork2017,santos2022investigating} which has demonstrated that repositories smaller than 10MB are unlikely to represent meaningful software projects.


In \textbf{Step 2}, we check whether repositories used the \textsc{GitHub Actions} \\ (\textsc{GHActions}) service. We focus on this CI service because it is the most widely used CI service in open-source projects, with a usage rate of 51.7\% \cite{GolzadehCIServer2022}. Furthermore, \textsc{GHActions} has become the most dominant CI/CD service only 18 months after its introduction, while other CI/CD services, especially Travis, have experienced decreasing adoption and increasing discontinuation rates \cite{DecanGitHubAction2022}.

Unlike other CI services, \textsc{GHActions} can define several workflows that can be responsible for executing tasks unrelated to CI. The documentation of 
\textsc{GHActions} states the following: \textit{``A workflow is an automated process composed of a series of jobs that gets executed when it's triggered by an event. Workflows are defined in YAML files and are stored in a .github/workflows directory at the root of the repository. A repository can also have multiple workflows''} \cite{GHActionGetStarted2013}. It was, therefore, necessary to identify which workflows were related to CI in \textbf{Step 3} and focus our analysis only on these workflows. To do so, we considered the following criteria:
\footnote{our scripts to identify CI-related workflows is publicly available on \url{https://github.com/jadsonjs/ci-workflows-trader}.}

\begin{sloppypar}
\begin{itemize}
\item  \textbf{Name Identification}: The workflow file contains ``CI'' in the name of the YAML file, for example: \texttt{``ci.yml''} or \texttt{``*-ci.yml''}, or contains ``CI'' in the  workflow name, for example: \texttt{``name: Node.js CI.''}
\item \textbf{Content Identification}: In this case, we identify workflows related to CI by the file content using three criteria: (i) checking if the workflow has the action \texttt{``action/checkout''}, as it is the most frequently used action in CI workflows \cite{DecanGitHubAction2022}, (ii) Check if the workflow has the event \texttt{``push''} or \texttt{``pull\_request''}, as they are the most frequent events~\cite{DecanGitHubAction2022}, and (iii) checking if the file has one of the 30 most common CI-related keywords in workflow file metadata. To find the most common CI-related keywords, we downloaded the content of all workflow files identified as CI workflows using the first criteria, tokenized it, and manually identified the 30 most common tokens related to CI. 
\end{itemize}
\end{sloppypar}

%
%
%

Using the criteria above, we identified a total of 19,142 workflows across 6,607 projects. The number of CI workflows per project has a range from 1 to 65. The first quartile (Q1) is 1, the third quartile (Q3) is 3, and the interquartile range (IQR) or median is 2. Notably, the distribution of workflows per project is skewed, with a small number of projects having a disproportionately high number of workflows.

In \textbf{Step 4}, we check if projects have been using CI for at least 2 years (last build - first build on \textsc{GHActions} >= 24 months) and if the last build happened in less than 6 months (last build <= 6 months). \textsc{GHActions} does not have the ``build'' concept as, for example, \textsc{TravisCI} does. We, therefore, consider a build to be a \texttt{Runner} of a CI workflow, which was considered in Step 3. In fact, \textsc{GHActions} define runners as: \textit{``processes on a server that run the workflow when it's triggered''}\cite{GHActionGetStarted2013}. We refer to \texttt{Runners} as builds for the sake of simplicity. We obtain a total of 2,488 projects with CI-related workflows. 

In \textbf{Step 5}, we collect information related to builds, issues, pull requests, commits, pull request comments, and code coverage within the period of the last two years of CI. Next, we filter for projects that have at least 100 builds, 100 issues, 100 pull requests, 100 commits, and 100 coverage entries. As for coverage, we focus on the \textsc{Coveralls}\footnote{https://docs.coveralls.io/api-introduction} and \textsc{CodeCov}\footnote{https://docs.codecov.com/reference/authorization} APIs, as they are two of the most popular coverage tools \cite{kavaler2019tool}. In total, we collected 871,104 pull request comments, 405,909 pull requests, 592,812 issues, 766,214 commits and 2,595,954 builds from 474 projects. This information was used to calculate CI metrics and identify the use of the evaluated CI practices in the projects.

\begin{figure}[!htbp]
  \centering
  \includegraphics[width=\linewidth]{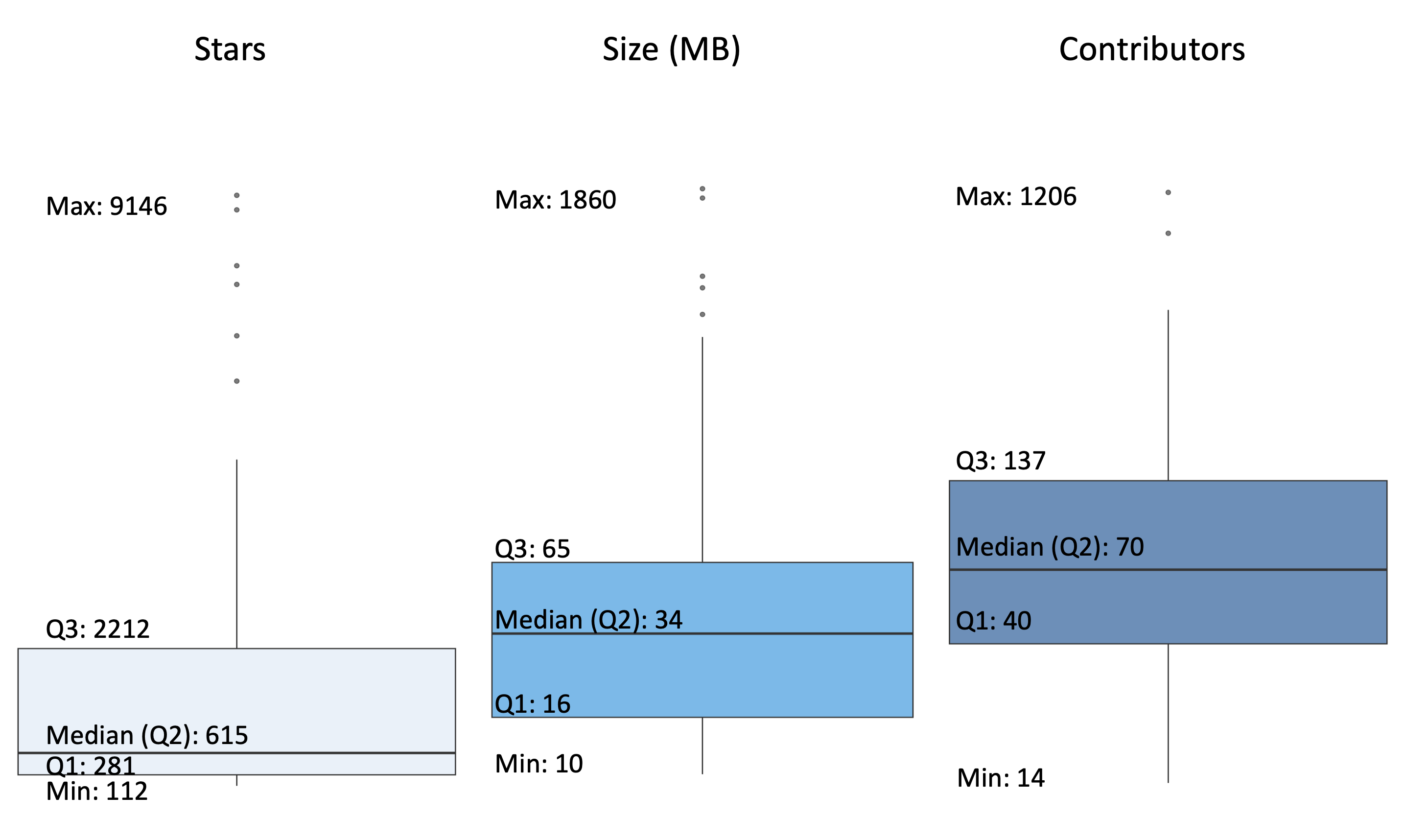}
  \caption{Contextual Factors of Projects in Our Database}
  \label{fig:projects_info}
\end{figure}

In \textbf{Step 6}, we filter for projects that have build information for each of the last 24 months (last build - 24 months), to select active projects and prevent projects with CI Theater \cite{CITheater2019}. This criterion ensured that the selected projects likely implemented CI practices rather than merely using a CI service (e.g., GitHub Actions) \cite{EliezioRevisaoSistematica2022}. Furthermore, since our analysis relied on build information, it was essential for these projects to include such data. This filter resulted in a \textbf{final set of 121 projects}. These projects together have a total sum of 73,004 PRs that were used to answer these research questions. We created a representative sample~\cite{SampleSize2021} of 382 randomly selected PRs (95\% confidence level and 5\% confidence interval) that contain at least one comment and were closed within the 24-month period during which data was collected for the study. It is important to note that the selected PRs were not stratified across the 121 projects, meaning they were chosen randomly without division or bias towards any specific project. Subsequently, we collected all the comments and reviews of these PRs, resulting in a total of 4,892 comments. PR comments generated by automated tools, such as bots were indeed included in our sample because we believe these comments may contain useful information generated by CI tools, and for this reason, they were not disregarded. Our random selection process obtained PR comments from 83 of the 121 final projects in our database. Figure \ref{fig:projects_info} presents some contextual factors of the projects in our database. 

We performed a Document Analysis \cite{bowen2009documentanalysis} following by Thematic Analysis~\cite{maguire2017doingThematicAnalysis} to identify needs and evidence of current monitoring practices (identifying metrics associated with these CI practices) in PR comments. We believe that Document Analysis combined with Thematic Analysis is the most appropriate choice for analyzing the data due to the large volume and richness of our dataset, which includes approximately 8GB of data including 871,104 PR comments. This extensive collection provides a unique opportunity to extract valuable insights into the developer processes across multiple projects. By systematically examining this large dataset, we can identify patterns, recurring themes, and behaviors related to how CI practices are monitored in these projects. While these comments are not exclusively about CI practices, they contain genuine discussions between developers, which may represent concerns about CI practices if they are pertinent to the project.

We extracted relevant text excerpts from the PR comments related to CI monitoring practices. We then conducted open coding by thoroughly examining the comments and identifying 417 codes. The coded data was analyzed to draw conclusions and insights regarding CI monitoring practices and the need for such monitoring. The Open Coding and Thematic Analysis was performed manually by the first author, without the support of automated tools, to ensure a comprehensive understanding of the data. The core steps of the process involved (1) initial open coding, where the comments were read in detail and codes were generated to represent key ideas; (2) searching for patterns, relationships, and similarities among the codes; and (3) the development of themes by grouping codes based on conceptual similarities. We identified nine key themes during our thematic analysis: (1) build duration, (2) code quality, (3) coverage, (4) hardware, (5) lead time for changes, (6) security, (7) build health and time to fix a broken build, (8) change failure rate and (9) deploy frequency. The themes were chosen to represent the overarching concepts that connect the codes within each theme. We group themes into 2 high-level themes: (i) ``Monitoring Opportunities'', which contains mentions of CI metrics but with no indication of monitoring those metrics (e.g. there are recurrent build failures, but no indication as to whether developers know the current or tolerated level of build failures) and (ii) ``Monitoring  Achievements'', which contains evidence that some CI metrics are monitored (e.g., automatic comments generated by the coverage analysis tools).

The extracted codes and themes were initially assigned by the first author. To reduce the probability of errors inherent in this type of subjective analysis, the second author independently applied the coding guide to label all examples. Finally, we calculated the agreement score between the two sets of labels using the Cohen Kappa statistical coefficient~\cite{cohen1968weighted}, which measures the agreement between two ``judges''.  We obtained a score of 0.68 (94.94\% of agreement), indicating ``Substantial Agreement''.

\subsection*{\textbf{RQ2: What is the perceived importance of monitoring CI practices?}}

\textbf{\textit{\underline{Motivation:}}} In light of the observed level of concern among developers regarding monitoring CI practices in RQ1, we delve deeper into this inquiry in RQ2 to directly solicit developers' opinions on the importance of monitoring CI practices by conducting a survey. We aim to address questions such as: Do developers prioritize monitoring their CI practices? Do they adhere to specific monitoring processes? Through RQ2, we seek  to understand developers' perceptions of CI practice monitoring and check whether they recognize its importance. Additionally, encourage discussion, knowledge sharing, and community building around best CI practices.

\hfill \break
\textbf{\textit{\underline{Approach:}}} From our dataset of 121 open-source projects from RQ1, we aim to recruit developers with knowledge and experience of CI practices to participate in our survey. In the \textbf{Step 7} of Figure~\ref{fig:approach_overview}, we select developers who had made significant contributions to these projects, as they were likely to have greater knowledge of the CI practices of their projects. We consider a developer to have made significant contributions if they had been involved with the project for a relatively long period of time and had more revisions in the version control system than the average developer~\cite{MiningRepositories2011}. We select developers with more commits than the average and who had committed at least 25\% of the 24-month study period (6 months). We identified 366 core developers, with an average of 3.02 commits, a median of 2, a maximum of 22, a minimum of 1, and a standard deviation of 3.57. We also randomly select an additional 134 developers to increase the pool of our survey participants, bringing the total number of potential participants to 500. Only developers with valid email were included. For example, we found that some emails were not valid, such as noreply@github.com or xxxxx@users.noreply.github.com, which were discarded.

In \textbf{Step 8} of Figure~\ref{fig:approach_overview}, to identify bug-related issues, we first collected labels associated with project issues. This approach is similar to Vasilescu et al. \cite{VasilescuQualityProdOutcomes2015}, where we manually check our projects to understand how they use labels to indicate the presence of bugs. We download all the labels from the 121 projects and checked their description to understand whether the label represented a bug. The resulting list of labels representing bug-related issues is: \textit{``defect'', ``error'', ``bug'', ``issue'', ``mistake'', ``incorrect'', ``fault'', ``flaw'', ``crash'', ``bugfix'' and ``regression.''} Similar to the work of Vasilescu et al.,~\cite{VasilescuQualityProdOutcomes2015}, we lowercased the labels and apply the Porter stemming algorithm \footnote{https://tartarus.org/martin/PorterStemmer/} to compare the bug-related labels. Finally, bug-related issues were analyzed to determine correlations between CI practices and the number of bugs in the projects. This data was used in an exploratory survey question to ascertain whether developers observed any correlations between CI practices and the incidence of bugs in the project.

To design our survey and improve participation, we adhere to the recommendations of Smith et al.'s work~\cite{ParticipationSurveys2013}. We (i) incentivized participants with the possibility of entering in a draw for gift cards; (ii) provided participants with an estimate of how long the survey would take; (iii) designed a relatively short and quick survey to complete; (iv) sent individual and personalized emails; and (v) invoked credibility by using the official titles and affiliations of researchers in the signature of the recruitment email as well as providing the contacts of the ethical board involved in the research.

The survey consisted of 18 questions, 11 of which were open-ended. To give respondents as much flexibility as possible, 8 questions were mandatory (7 of which were close-ended while 1 was open-ended), and 10 questions were optional (all open-ended). Participation was voluntary and the estimated time to complete the survey was 10 to 15 minutes. A total of 500 emails were sent out over a period of 45 days (January 30 to March 13, 2023). We received 28 responses, all of which were considered valid as the participants completed the mandatory questions. \textbf{Overall, our survey achieved a response rate of 5.6\%}. Across all 18 questions, participants answered an average of 15.25 questions, with a minimum of 8 (a single occurrence) and a maximum of 18 (nine occurrences). In relation to the number of responses received for each optional question, the optional question with the highest number of responses received 26 answers, while the least answered question received 15 responses with an average of 21 answers.

We generated a customized survey for each of the 121 final projects, detailing the evolution of the project's specific CI practices. To achieve this, we calculated the values of seven CI metrics associated with these practices, which were presented to the developers who participated in the survey. Table\ref{tab:ci-metrics} presents the CI metrics used in the survey, along with their definitions and the related CI practices we want to evaluate.

\begin{table}
\caption{ Continuous Integration Practices, Metrics and Definitions}
\label{tab:ci-metrics}
\begin{tabular}{lll}
\hline
\textbf{CI Practice}               & \textbf{CI Metric}            & \textbf{CI Metric Definition}\\
\hline                                                                                                                                                                                                                                                                                                                                                                                                                                                                  \\
Commit Code Frequently \cite{Duvall2007}          & Commit Per Weekday \cite{CITheater2019}        & \textit{\begin{tabular}[c]{@{}l@{}}Mean of the absolute number of  \\ commits  according to the weekday \\ of the analyzed period.\end{tabular}}                   \\ \hline
\begin{tabular}[c]{@{}l@{}} Write Automated \\ Developer Tests \end{tabular} \cite{Duvall2007} & Coverage \cite{CITheater2019}                  & \textit{\begin{tabular}[c]{@{}l@{}}It measures which parts of a  \\ program are executed \\ when running the tests.\end{tabular}}                                    \\ \hline
Keep the Build Fast \cite{fowler2006continuous}            & Build Duration \cite{CITheater2019}            & \textit{ \begin{tabular}[c]{@{}l@{}}It measures the duration \\ of the build. The time between \\ the start and end of the build.\end{tabular} }                                                                                                                  \\ \hline
Automate the Build \cite{fowler2006continuous}             & Build Activity \cite{STAHL201448}            & \textit{\begin{tabular}[c]{@{}l@{}}It is a unit interval \\ (i.e., a closed interval {[}0,1{]}) \\ representing the rate of builds \\ across days.\end{tabular}} \\ \hline
\begin{tabular}[c]{@{}l@{}}  Don't Commit \\ Broken Code \end{tabular} \cite{Duvall2007,fowler2006continuous}       & Build Health \cite{EliezioRevisaoSistematica2022}              & \textit{\begin{tabular}[c]{@{}l@{}}It is a unit interval representing \\ the rate of build failures \\ across days.\end{tabular}}                                   \\ \hline
\begin{tabular}[c]{@{}l@{}} Fix Broken Builds \\ Immediately \end{tabular} \cite{Duvall2007,fowler2006continuous}  & \begin{tabular}[c]{@{}l@{}} Time to Fix \\ a Broken Build \cite{CITheater2019} \end{tabular} & \textit{\begin{tabular}[c]{@{}l@{}}It consists of the median time  \\ in a period that builds remained \\ broken.\end{tabular}}                                      \\ \hline
Conduct Code Reviews \cite{mcintosh2016empirical}        & Comments Per Change \cite{thompson2017large,staahl2013experienced}       & \textit{\begin{tabular}[c]{@{}l@{}}Mean of the number of comments \\ grouped by changes. \end{tabular}}                                                            \\
\end{tabular}
\end{table}

Before presenting the evolution charts for the seven CI metrics we quantified for the projects, we outline the metrics and describe how we measured them for the purpose of the survey.

To apply the survey, we use definitions of CI maturity model from the gray literature \cite{DeliveryMaturityModel2013}. From this model, we use just the two categories most related to the practices collected from the projects: (i) Build \& Deploy; and (ii) Test \& Verification. The final definition of the five levels of CI maturity used in the survey was:

\begin{itemize}

\item{\textbf{Base}: The deployment process is manual or semi-manual with some parts scripted and rudimentarily documented in some way. The team will typically practice unit-testing and have one or more dedicated test environments.}

\item{\textbf{Beginner}: Introduces frequent polling builds for faster feedback and build artifacts are archived for easier dependency management. The team starts to investigate ways of gradually automating the existing manual integration testing.}

\item{\textbf{Intermediate}: Builds are typically triggered from the source control system on each commit, and the deployment process is standardized over all environments. At this level you will most likely start to look at gradually automating parts of the acceptance testing.}

\item{\textbf{Advanced}: Techniques for zero downtime deploys can be important to include in the automated process. Techniques of database migrations can be used to completely avoid manual routines for database updates. Fully automatic acceptance tests and maybe also generating structured acceptance criteria directly from requirements.}

\item{\textbf{Expert}: Every commit can potentially make it all the way to production automatically. Infrastructure as code making the build process produces not only deployment artifacts but complete virtual machines. The organization, culture and tooling have reached a certain maturity level and feedback on relevant business metrics is fast and accessible.}

\end{itemize}

\begin{figure}[!htbp]
  \centering
  \includegraphics[width=\linewidth]{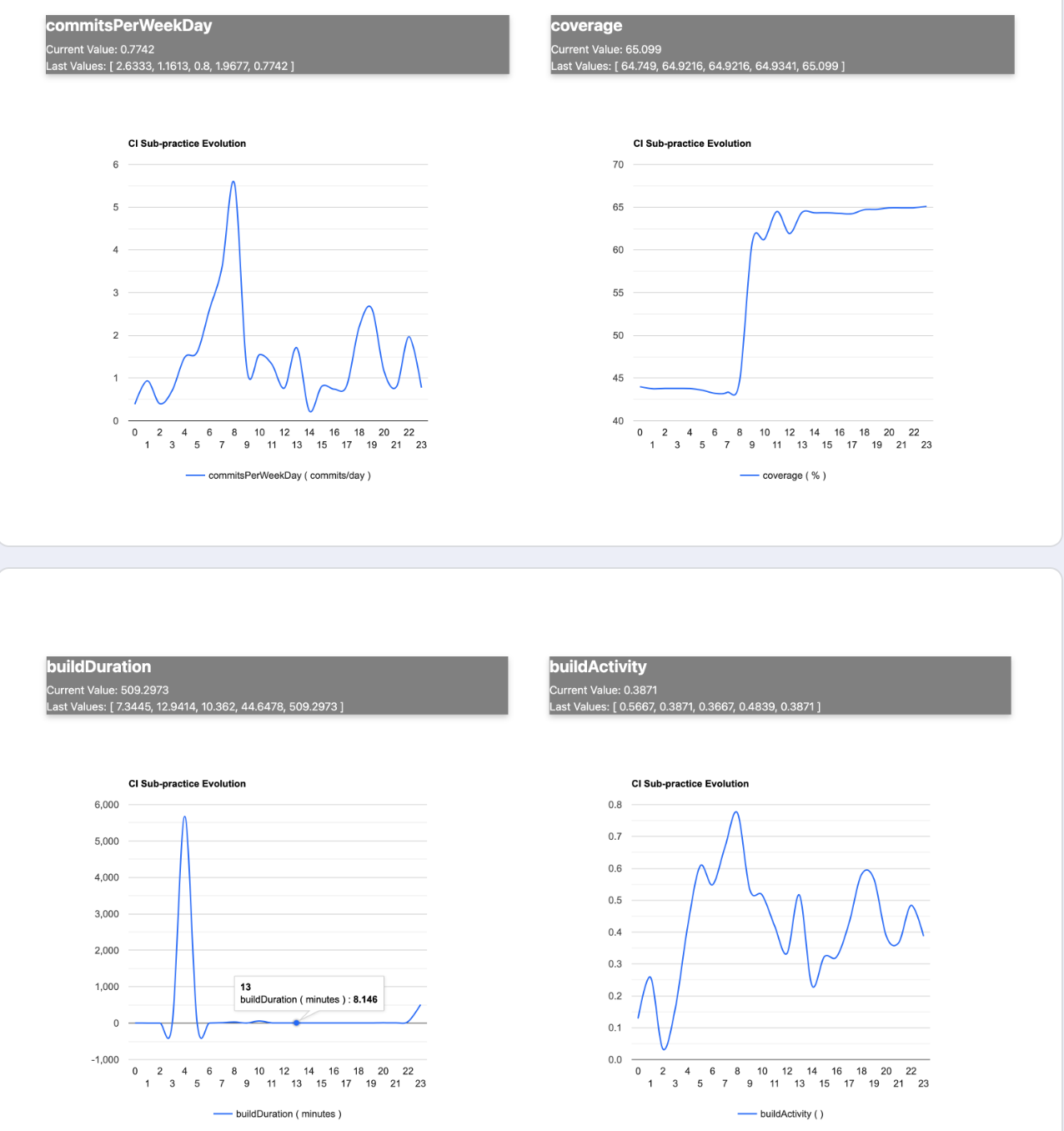}
  \caption{Partial evolution charts for the seven CI metrics}
  \label{fig:practices_evolution}
\end{figure}

\begin{figure}[!htbp]
  \centering
  \includegraphics[width=0.8\linewidth]{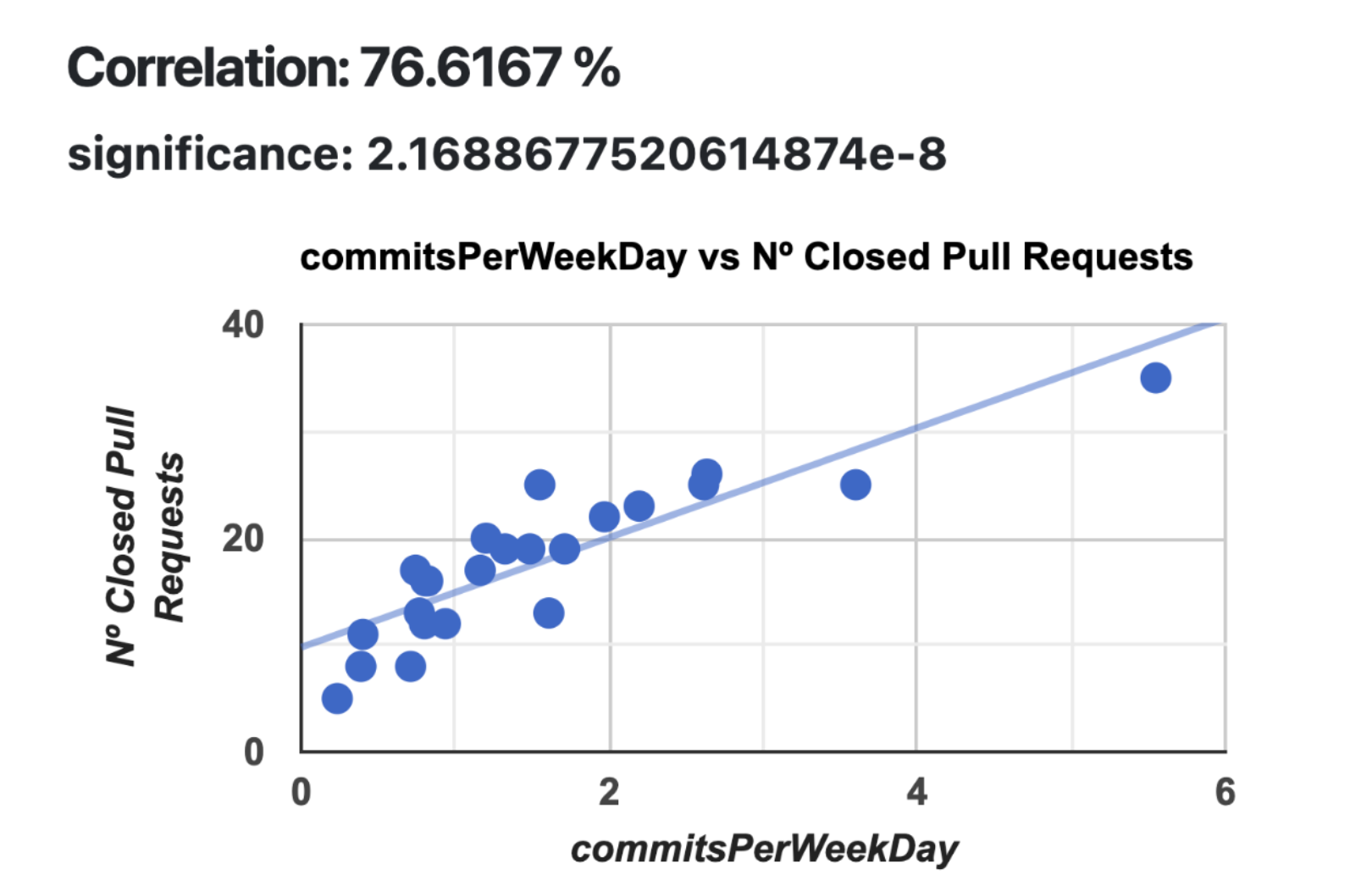}
  \caption{Partial correlations between the seven CI metrics and quality attributes}
  \label{fig:practice_correlation}
\end{figure}

The survey was divided into four sections:

\begin{enumerate}
\item  \textbf{Introduction}: In this section, we provided participants with the survey context. We explained the purpose of the research, the estimated response time, the data collected would be used, and the compensation that would be offered to those who completed the survey.

\item \textbf{Personal information}: In this section, we asked for information about the participants' experience in software development and their involvement in the project's adoption or configuration of CI. 

\item \textbf{Perception of CI practices}: In this set of questions, we asked participants about the current state of CI in their projects. This included the level of CI maturity, what metrics are under monitoring and how often, and which tools are used.

\item \textbf{Evaluation of CI practice}: In the last part of the survey, we presented evolution charts for the seven CI metrics that we quantified for the project in question, as shown in Figure \ref{fig:practices_evolution}. We asked participants to rate the importance of these metrics to the project and whether they deemed it important to measure the evolution of these metrics over time. We also asked if the values presented for these metrics were surprising or worrying in any way.

For some projects, we presented correlations between CI metrics and the number of closed pull requests (an indicator of effort) and the number of bug-related issues (an indicator of quality), as shown Figure \ref{fig:practice_correlation}. We requested participants to provide their comments regarding the correlations shown. Only significant correlations, above 50\%, were shown. Otherwise, these correlations were omitted from participants.

Finally, we asked if there was any missing information that our participants would like to know of or if they would like to receive regular updates. We also requested participants to provide their email should they wish to participate in the prize draw. We provide a complete copy of our questionnaire and dataset in our public repository.\footnote{https://doi.org/10.5281/zenodo.14569025}

\end{enumerate}

We compile all survey results into a single spreadsheet. We show the results from the closed-ended responses in a graphical format, while we perform a Thematic Analysis of the open-ended responses, using quotes to illustrate the themes whenever necessary. We group the responses into seven themes, which are listed below:



\begin{itemize}
\item \textbf{Perceptions of CI maturity level}:  This theme groups together the responses regarding developers' perceptions of the maturity level of their projects in relation to CI practices.
\item \textbf{Monitoring CI practices}: This theme groups the findings on how developers currently monitor CI practices.
\item \textbf{Currently used CI Tools}: This theme summarizes the findings on the tools currently used to monitor CI practices.
\item \textbf{Evaluation of CI practices}: In this theme, we describe the respondents' perceptions of the seven CI practices collected in their project context. What developers think about CI practices and whether they are important to the project from their perspective.
\item \textbf{Impact of CI practices}: We provide developers' perceptions of the impact of CI practices on their projects.
\item \textbf{Correlations with Quality Attributes}: We provide developers' perceptions of how useful the correlation between CI practices and Quality Attributes is in their projects.
\item \textbf{Considerations about Monitoring}: We listed the considerations expressed by respondents for monitoring CI practices.
\end{itemize} 

The steps involved in our study follow the guidelines from Linaker et al.~\cite{linaker2015guidelines}, which were: defining the research questions, defining the target audience,  creating the survey, evaluating it on a subset of the sample population, conducting the survey, analyzing the results and drawing conclusions. We created a first draft of the survey and went through five rounds of revision with different participants drawn from our professional network (i.e., not the core developers yet), until we reached a refined version of the survey. We sent this refined version to 20\% ($\frac{100}{500}$) of developers and awaited for the first responses (15 days). Once 6 responses were obtained, we made small adjustments to the questions, improving the clarity of our survey. We then sent this final version to the remaining 80\% ($\frac{400}{500}$) of developers.

\subsection*{\textbf{RQ3: Which monitoring features are currently supported by existing CI tools?}}

\textbf{\textit{\underline{Motivation:}}} In light of the interest in monitoring CI practices, as observed in RQ1 and corroborated in RQ2, we investigate the current features available for monitoring CI practices. This research question aims to enhance developers' awareness of the existing monitoring capabilities of CI services and third-party tools and identify any missing capabilities that could be potentially developed or enhanced in the future, identifying any gaps or missing features in current monitoring tools. The findings of this research question can inform future research and development efforts in the field of CI monitoring. Researchers and tool developers can prioritize areas for improvement based on the identified needs.


\hfill \break
\textbf{\textit{\underline{Approach:}}} In RQ3, we conduct a Gray Literature Review (GLR) \cite{garousi2020benefitting} to compile the existing monitoring capabilities of existing CI tools and services. Since this method enables faster acquisition of research results in practical contexts, it more accurately reflects the current capabilities of CI services and third-party tools in monitoring CI practices. To systematically collate existing monitoring capabilities, we conduct a Document Analysis \cite{bowen2009documentanalysis} on the official documentation of main CI services. In our analysis, we focused on the seven most popular CI services, covering 99\% of all observed instances of CI usage \cite{GolzadehCIServer2022}: \textsc{TravisCI}, \textsc{GHActions}, \textsc{CircleCI}, \textsc{AppVeyor}, \textsc{Azure}, \textsc{GitLab} and \textsc{Jenkins}. 

As an initial step, we conduct a search on the official documentation of each CI service mentioned above, as this is the most intuitive thing a developer does when he wants to learn about these CI services. Furthermore, these documents can be classified according to \cite{garousi2019guidelines} as ``2nd tier Gray Literature'', which provides ``Moderate credibility/Moderate outlet control''. The use of a specific data source also allows the study to be reproducible. Subsequently, we meticulously examined the sections within the documentation that could be directly correlated with the concept of monitoring. When the official documentation had links to third-party documentation, we expanded our examination to include the third-party documentation, using a snowballing process. We discover monitoring evidence in 37 documents, with 27 being primary documents and 10 being snowballing documents. 

As an additional step, to increase the number of documents, we use query strings on google.com, using incognito windows to collect official documentation from these services. We used the first ten official documentation web pages in the search results as input for our Document Analysis. We searched within the first 100 results.

To collect the official documentation from the CI services, we used the following query string on \url{google.com}, using incognito windows:

\begin{itemize}
\item  \textit{ (<CI Service Name>\footnote{<CI Service Name> = ``Travis'', ``github actions'', ``Jenkins'', ``GitLab'', ``CircleCI'', ``AppVeyor'', ``Azure''}) AND (``continuous integration'') AND (practices OR metrics) AND monitoring }
\end{itemize}

In this step, we ignore third-party documentation and sponsored links to ensure the trustworthiness of the data \cite{morgan2022conductingDocumentAnalysis}. We examine additional 33 documents with this methodology. In Table \ref{tab:ci-servers-oficial-documents} and Table \ref{tab:ci-servers-oficial-documents-continuation}, we list the documents found during the two steps of research in which the Thematic Analysis was applied. We provide Gray Literature URLs and store all searched and collected data in an external database\footnote{https://doi.org/10.5281/zenodo.14569025} for later consultation to increase study's replicability \cite{KAMEI2021106609}.

\begin{table*}
\caption{Official documents analyzed from CI services}
\label{tab:ci-servers-oficial-documents}
\begin{tabular}{llll}
\hline
\textbf{CI service}  & \textbf{Method} & \textbf{Scope} & \textbf{Analyzed Documents}        \\ \hline

 \multirow{13}{*}{GITHUB ACTIONS}     & \cellcolor[HTML]{ECF4FF}manual & \cellcolor[HTML]{9AFF99}primary & \begin{tabular}[c]{@{}l@{}} 1 - https://docs.github.com/en/actions/\\monitoring-and-troubleshooting-workflows/\\about-monitoring-and-troubleshooting \end{tabular} \\ 
                                      & \cellcolor[HTML]{ECF4FF}manual & \cellcolor[HTML]{9AFF99}primary & \begin{tabular}[c]{@{}l@{}}2 - https://docs.github.com/en/actions/\\monitoring-and-troubleshooting-workflows/\\using-the-visualization-graph\end{tabular} \\ 
                                      & \cellcolor[HTML]{ECF4FF}manual & \cellcolor[HTML]{9AFF99}primary & \begin{tabular}[c]{@{}l@{}}3 - https://docs.github.com/en/actions/\\monitoring-and-troubleshooting-workflows/\\adding-a-workflow-status-badge\end{tabular} \\ 
                                      & \cellcolor[HTML]{ECF4FF}manual & \cellcolor[HTML]{9AFF99}primary & \begin{tabular}[c]{@{}l@{}}4 - https://docs.github.com/en/actions/\\monitoring-and-troubleshooting-workflows/\\using-workflow-run-logs\end{tabular} \\ 
                                      & \cellcolor[HTML]{FFFFC7}query & \cellcolor[HTML]{9AFF99}primary & \begin{tabular}[c]{@{}l@{}}5 - https://github.com/marketplace/\\actions/datadog-actions-metrics \end{tabular} \\ 
                                      & \cellcolor[HTML]{FFFFC7}query & \cellcolor[HTML]{9AFF99}primary & \begin{tabular}[c]{@{}l@{}}6 - https://github.com/marketplace/\\category/monitoring \end{tabular} \\ 
                                      & \cellcolor[HTML]{FFFFC7}query & \cellcolor[HTML]{9AFF99}primary & \begin{tabular}[c]{@{}l@{}}7 - https://github.com/marketplace/\\actions/workflow-telemetry \end{tabular} \\ 
                                      & \cellcolor[HTML]{FFFFC7}query & \cellcolor[HTML]{9AFF99}primary & \begin{tabular}[c]{@{}l@{}}8 - https://github.com/topics/\\continuous-integration?l=shell\&o=asc\&s=forks \end{tabular} \\ 
                                      & \cellcolor[HTML]{FFFFC7}query & \cellcolor[HTML]{9AFF99}primary & \begin{tabular}[c]{@{}l@{}}9 - https://resources.github.com/\\devops/tools/automation/actions/ \end{tabular} \\ 
                                      & \cellcolor[HTML]{FFFFC7}query & \cellcolor[HTML]{9AFF99}primary & \begin{tabular}[c]{@{}l@{}}10 - https://github.com/topics/github-action \end{tabular}\\ 
                                      & \cellcolor[HTML]{FFFFC7}query & \cellcolor[HTML]{9AFF99}primary & \begin{tabular}[c]{@{}l@{}}11 - https://github.com/marketplace/\\actions/lighthouse-ci-action \end{tabular} \\ 
                                      & \cellcolor[HTML]{FFFFC7}query & \cellcolor[HTML]{9AFF99}primary & \begin{tabular}[c]{@{}l@{}}12 - https://github.com/iterative/setup-cml \end{tabular} \\ 
                                      & \cellcolor[HTML]{FFFFC7}query & \cellcolor[HTML]{9AFF99}primary & \begin{tabular}[c]{@{}l@{}}13 - https://resources.github.com/ci-cd/ \end{tabular}  \\ 
                                      \hline

 \multirow{14}{*}{GITLAB}     & \cellcolor[HTML]{ECF4FF}manual & \cellcolor[HTML]{9AFF99}primary & \begin{tabular}[c]{@{}l@{}} 1 - https://docs.gitlab.com/\\ee/user/group/\\value\_stream\_analytics/ \end{tabular} \\ 
 & \cellcolor[HTML]{ECF4FF}manual & \cellcolor[HTML]{9AFF99}primary & \begin{tabular}[c]{@{}l@{}} 2 - https://docs.gitlab.com/\\ee/user/analytics/dora\_metrics.html \end{tabular}  \\
                              & \cellcolor[HTML]{ECF4FF}manual & \cellcolor[HTML]{9AFF99}primary & \begin{tabular}[c]{@{}l@{}} 3 - https://docs.gitlab.com/\\ee/user/analytics/value\_streams\_dashboard.html \end{tabular}  \\
                              & \cellcolor[HTML]{ECF4FF}manual & \cellcolor[HTML]{9AFF99}primary & \begin{tabular}[c]{@{}l@{}} 4 - https://docs.gitlab.com/\\ee/user/analytics/ci\_cd\_analytics.html \end{tabular}  \\
                              & \cellcolor[HTML]{FFFFC7}query  & \cellcolor[HTML]{9AFF99}primary & \begin{tabular}[c]{@{}l@{}} 5 - https://about.gitlab.com/\\topics/ci-cd/continuous-integration-metrics/ \end{tabular}  \\
                              & \cellcolor[HTML]{FFFFC7}query  & \cellcolor[HTML]{9AFF99}primary & \begin{tabular}[c]{@{}l@{}} 6 - https://docs.gitlab.com/\\ee/ci/introduction/\end{tabular}  \\
                              & \cellcolor[HTML]{FFFFC7}query  & \cellcolor[HTML]{9AFF99}primary & \begin{tabular}[c]{@{}l@{}} 7 - https://docs.gitlab.com/\\ee/ci/ \end{tabular}  \\
                              & \cellcolor[HTML]{FFFFC7}query  & \cellcolor[HTML]{9AFF99}primary & \begin{tabular}[c]{@{}l@{}} 8 - https://about.gitlab.com/\\blog/2022/02/03/\\how-to-keep-up-with-ci-cd-best-practices/ \end{tabular}  \\
                              & \cellcolor[HTML]{FFFFC7}query  & \cellcolor[HTML]{9AFF99}primary & \begin{tabular}[c]{@{}l@{}} 9 - https://about.gitlab.com/\\features/continuous-integration/ \end{tabular}  \\
                              & \cellcolor[HTML]{FFFFC7}query  & \cellcolor[HTML]{9AFF99}primary & \begin{tabular}[c]{@{}l@{}} 10 - https://about.gitlab.com/blog/2022/06/14/\\observability-vs-monitoring-in-devops/ \end{tabular} \\
                              & \cellcolor[HTML]{FFFFC7}query  & \cellcolor[HTML]{9AFF99}primary & \begin{tabular}[c]{@{}l@{}} 11 - https://about.gitlab.com/\\stages-devops-lifecycle/\\continuous-delivery/ \end{tabular} \\
                              & \cellcolor[HTML]{FFFFC7}query  & \cellcolor[HTML]{9AFF99}primary & \begin{tabular}[c]{@{}l@{}} 12 - https://about.gitlab.com/\\stages-devops-lifecycle/monitor/ \end{tabular} \\
                              & \cellcolor[HTML]{FFFFC7}query  & \cellcolor[HTML]{9AFF99}primary & \begin{tabular}[c]{@{}l@{}} 13 - https://about.gitlab.com/\\blog/2020/12/17/cd-solution-overview/ \end{tabular} \\
                              & \cellcolor[HTML]{FFFFC7}query  & \cellcolor[HTML]{9AFF99}primary & \begin{tabular}[c]{@{}l@{}} 14 - https://about.gitlab.com/handbook/\\marketing/brand-and-product-marketing/ \\ product-and-solution-marketing/usecase-gtm/ci/ \\ \end{tabular} \\
                             \hline

\multirow{15}{*}{TRAVIS CI}           & \cellcolor[HTML]{ECF4FF}manual & \cellcolor[HTML]{9AFF99}primary & \begin{tabular}[c]{@{}l@{}} 1 - https://docs.travis-ci.com/user/notifications/ \end{tabular}  \\
                                      & \cellcolor[HTML]{ECF4FF}manual & \cellcolor[HTML]{9AFF99}primary & \begin{tabular}[c]{@{}l@{}} 2 - https://docs.travis-ci.com/user/status-images/ \end{tabular}  \\
                                      & \cellcolor[HTML]{ECF4FF}manual & \cellcolor[HTML]{9AFF99}primary & \begin{tabular}[c]{@{}l@{}} 3 - https://docs.travis-ci.com/user/code-climate/ \end{tabular}  \\
                                      & \cellcolor[HTML]{ECF4FF}manual & \cellcolor[HTML]{9AFF99}primary & \begin{tabular}[c]{@{}l@{}} 4 - https://docs.travis-ci.com/user/deepsource/ \end{tabular} \\
                                      & \cellcolor[HTML]{ECF4FF}manual & \cellcolor[HTML]{9AFF99}primary & \begin{tabular}[c]{@{}l@{}} 5 - https://docs.travis-ci.com/user/coveralls/ \end{tabular} \\
                                      & \cellcolor[HTML]{ECF4FF}manual & \cellcolor[HTML]{9AFF99}primary & \begin{tabular}[c]{@{}l@{}} 6 - https://docs.travis-ci.com/user/coverity-scan/ \end{tabular} \\
                                      & \cellcolor[HTML]{ECF4FF}manual & \cellcolor[HTML]{9AFF99}primary & \begin{tabular}[c]{@{}l@{}} 7 - https://docs.travis-ci.com/user/sonarcloud/ \end{tabular} \\
                                      & \cellcolor[HTML]{ECF4FF}manual & \cellcolor[HTML]{9AFF99}primary & \begin{tabular}[c]{@{}l@{}} 8 - https://docs.travis-ci.com/user/sourceclear/ \end{tabular} \\
                                      & \cellcolor[HTML]{ECF4FF}manual & \cellcolor[HTML]{9AFF99}primary & \begin{tabular}[c]{@{}l@{}} 9 - https://docs.travis-ci.com/user/cc-menu/ \end{tabular} \\
                                      & \cellcolor[HTML]{ECF4FF}manual & \cellcolor[HTML]{FFCCC9}snowballing & \begin{tabular}[c]{@{}l@{}} 10 - http://marcells.github.io/node-build-monitor/ \end{tabular} \\
                                      & \cellcolor[HTML]{ECF4FF}manual & \cellcolor[HTML]{FFCCC9}snowballing & \begin{tabular}[c]{@{}l@{}} 11 - https://github.com/ahsayde/ci-dashboard \end{tabular} \\
                                      & \cellcolor[HTML]{ECF4FF}manual & \cellcolor[HTML]{FFCCC9}snowballing & \begin{tabular}[c]{@{}l@{}} 12 - https://buildmonitor.io/ \end{tabular} \\
                                      & \cellcolor[HTML]{ECF4FF}manual & \cellcolor[HTML]{FFCCC9}snowballing & \begin{tabular}[c]{@{}l@{}} 13 - https://meercode.io/ \end{tabular} \\
                                      & \cellcolor[HTML]{ECF4FF}manual & \cellcolor[HTML]{FFCCC9}snowballing & \begin{tabular}[c]{@{}l@{}} 14 - https://buildpulse.io/ \end{tabular} \\
                                      & \cellcolor[HTML]{ECF4FF}manual & \cellcolor[HTML]{FFCCC9}snowballing & \begin{tabular}[c]{@{}l@{}} 15 - http://fdietz.github.io/team\_dashboard/ \end{tabular} \\
                                      \hline

\end{tabular}
\end{table*}

\begin{table*}
\caption{Official documents analyzed from CI services (continuation) }
\label{tab:ci-servers-oficial-documents-continuation}
\begin{tabular}{llll}
\hline
\textbf{CI service}  & \textbf{Method} & \textbf{Scope} & \textbf{Analyzed Documents}        \\ \hline

\multirow{11}{*}{JENKINS}             & \cellcolor[HTML]{ECF4FF}manual & \cellcolor[HTML]{9AFF99}primary & \begin{tabular}[c]{@{}l@{}} 1 - https://www.jenkins.io/blog/2016/07/18/\\pipeline-notifications/ \end{tabular}  \\
                                      & \cellcolor[HTML]{ECF4FF}manual & \cellcolor[HTML]{FFCCC9}snowballing & \begin{tabular}[c]{@{}l@{}} 2 - https://plugins.jenkins.io/\\pipeline-stage-view/ \end{tabular}  \\
                                      & \cellcolor[HTML]{ECF4FF}manual & \cellcolor[HTML]{9AFF99}primary & \begin{tabular}[c]{@{}l@{}} 3 - https://www.jenkins.io/doc/book/\\pipeline/getting-started/ \end{tabular}  \\
                                      & \cellcolor[HTML]{ECF4FF}manual & \cellcolor[HTML]{9AFF99}primary & \begin{tabular}[c]{@{}l@{}} 4 - https://www.jenkins.io/doc/book/\\pipeline/running-pipelines/ \end{tabular}  \\
                                      & \cellcolor[HTML]{ECF4FF}manual  & \cellcolor[HTML]{9AFF99}primary & \begin{tabular}[c]{@{}l@{}} 5 - https://www.jenkins.io/doc/book/\\pipeline/pipeline-as-code/ \end{tabular}  \\
                                      & \cellcolor[HTML]{ECF4FF}manual  & \cellcolor[HTML]{FFCCC9}snowballing & \begin{tabular}[c]{@{}l@{}} 6 - https://github.com/newrelic/\\nr-jenkins-plugin\#readme \end{tabular}  \\
                                      & \cellcolor[HTML]{ECF4FF}manual  & \cellcolor[HTML]{9AFF99}primary & \begin{tabular}[c]{@{}l@{}} 7 - https://plugins.jenkins.io/datadog/ \end{tabular} \\
                                      & \cellcolor[HTML]{ECF4FF}manual  & \cellcolor[HTML]{FFCCC9}snowballing & \begin{tabular}[c]{@{}l@{}} 8 - https://github.com/jenkinsci/metrics-plugin \end{tabular} \\
                                      & \cellcolor[HTML]{ECF4FF}manual  & \cellcolor[HTML]{FFCCC9}snowballing & \begin{tabular}[c]{@{}l@{}} 9 - https://github.com/jenkinsci/\\prometheus-plugin/blob/master/docs\\/metrics/index.md \end{tabular} \\
                                      & \cellcolor[HTML]{FFFFC7}query  & \cellcolor[HTML]{9AFF99}primary & \begin{tabular}[c]{@{}l@{}} 10 - https://plugins.jenkins.io/\\instance-identity/dependencies/ \end{tabular} \\
                                      & \cellcolor[HTML]{FFFFC7}query  & \cellcolor[HTML]{9AFF99}primary & \begin{tabular}[c]{@{}l@{}} 11 - https://www.jenkins.io/\\doc/pipeline/steps/ \end{tabular} \\
                                      \hline

\multirow{12}{*}{CIRCLE CI}           & \cellcolor[HTML]{ECF4FF}manual & \cellcolor[HTML]{9AFF99}primary & \begin{tabular}[c]{@{}l@{}} 1 - https://circleci.com/docs/insights/ \end{tabular}  \\
                                      & \cellcolor[HTML]{ECF4FF}manual & \cellcolor[HTML]{9AFF99}primary & \begin{tabular}[c]{@{}l@{}} 2 - https://circleci.com/docs/insights-snapshot-badge/ \end{tabular}  \\
                                      & \cellcolor[HTML]{ECF4FF}manual & \cellcolor[HTML]{9AFF99}primary & \begin{tabular}[c]{@{}l@{}} 3 - https://circleci.com/docs/insights-glossary/ \end{tabular}  \\
                                      & \cellcolor[HTML]{FFFFC7}query & \cellcolor[HTML]{9AFF99}primary & \begin{tabular}[c]{@{}l@{}} 4 - https://circleci.com/continuous-integration/ \end{tabular}  \\
                                      & \cellcolor[HTML]{FFFFC7}query  & \cellcolor[HTML]{9AFF99}primary & \begin{tabular}[c]{@{}l@{}} 5 - https://circleci.com/blog/engineering-metrics/ \end{tabular}  \\
                                      & \cellcolor[HTML]{FFFFC7}query  & \cellcolor[HTML]{9AFF99}primary & \begin{tabular}[c]{@{}l@{}} 6 - https://circleci.com/blog/ci-cd-code-quality-metrics/ \end{tabular}  \\
                                      & \cellcolor[HTML]{FFFFC7}query  & \cellcolor[HTML]{9AFF99}primary & \begin{tabular}[c]{@{}l@{}} 7 - https://circleci.com/blog/\\a-brief-history-of-devops-part-iv-continuous-delivery\\-and-continuous-deployment/ \end{tabular}  \\
                                      & \cellcolor[HTML]{FFFFC7}query  & \cellcolor[HTML]{9AFF99}primary & \begin{tabular}[c]{@{}l@{}} 8 - https://circleci.com/blog/\\benefits-of-scheduling-ci-pipelines/ \end{tabular}  \\
                                      & \cellcolor[HTML]{FFFFC7}query  & \cellcolor[HTML]{9AFF99}primary & \begin{tabular}[c]{@{}l@{}} 9 - https://circleci.com/blog/observability-vs-monitoring/ \end{tabular}  \\
                                      & \cellcolor[HTML]{FFFFC7}query  & \cellcolor[HTML]{9AFF99}primary & \begin{tabular}[c]{@{}l@{}} 10 - https://circleci.com/blog/\\how-to-measure-devops-success-4-key-metrics/ \end{tabular} \\
                                      & \cellcolor[HTML]{FFFFC7}query  & \cellcolor[HTML]{9AFF99}primary & \begin{tabular}[c]{@{}l@{}} 11 - https://circleci.com/docs/overview/ \end{tabular} \\
                                      & \cellcolor[HTML]{FFFFC7}query  & \cellcolor[HTML]{9AFF99}primary & \begin{tabular}[c]{@{}l@{}} 12 - https://circleci.com/landing-pages/assets/\\2022-state-of-software-delivery-report.pdf \end{tabular} \\
                                      \hline

\multirow{2}{*}{APPVEYOR}             & \cellcolor[HTML]{ECF4FF}manual & \cellcolor[HTML]{9AFF99}primary & \begin{tabular}[c]{@{}l@{}} 1 - https://www.appveyor.com/docs/notifications/ \end{tabular}  \\
                                      & \cellcolor[HTML]{FFFFC7}query  & \cellcolor[HTML]{9AFF99}primary & \begin{tabular}[c]{@{}l@{}} 2 - https://www.appveyor.com/ \end{tabular}  \\
                                      \hline

\multirow{3}{*}{AZURE}               & \cellcolor[HTML]{ECF4FF}manual & \cellcolor[HTML]{9AFF99}primary & \begin{tabular}[c]{@{}l@{}} 1 - https://learn.microsoft.com/pt-br/\\azure/devops/pipelines/get-started/\\what-is-azure-pipelines?view=azure-devops \end{tabular}  \\
                                      & \cellcolor[HTML]{FFFFC7}query  & \cellcolor[HTML]{9AFF99}primary & \begin{tabular}[c]{@{}l@{}} 2 - https://azure.microsoft.com/pt-br/blog/\\7-best-practices-for-continuous-\\monitoring-with-azure-monitor/ \end{tabular}  \\
                                      & \cellcolor[HTML]{FFFFC7}query  & \cellcolor[HTML]{9AFF99}primary & \begin{tabular}[c]{@{}l@{}} 3 - https://learn.microsoft.com/en-us/azure\\/architecture/example-scenario/apps/\\devops-dotnet-baseline \end{tabular}  \\
                                      \hline  

\end{tabular}
\end{table*}

In the third step of RQ3, we decided to include the analysis of third-party tools, as these tools may have features not present in the main CI services. We collect all references to third-party tools cited in the official CI services documentation obtained from the previous two steps. Subsequently, we conduct a Document Analysis on the documentation of these third-party tools. In this step, we identify 31 mentions of third-party tools and conduct a complementary Document Analysis on 41 documents related to these tools. Similar to the first steps, we provided Gray Literature URLs and stored all the data searched and collected in an external database\footnote{https://doi.org/10.5281/zenodo.14569025} for later consultation to enhance the replicability of the study. In Table \ref{tab:third-party-oficial-documents} and Table \ref{tab:third-party-oficial-documents-2}, we list all the documents analyzed.

\begin{table*}
\caption{Official documents analyzed from third-party tools}
\label{tab:third-party-oficial-documents}
\begin{tabular}{lllll}
\hline
\textbf{\#} & \textbf{third-party tool} & \textbf{Method} & \textbf{Scope} & \textbf{Analyzed Documents}        \\ \hline


\multirow{7}{*}{1} & \multirow{7}{*}{DATADOG}  
    & \cellcolor[HTML]{ECF4FF}manual & \cellcolor[HTML]{FFCCC9}snowballing & \begin{tabular}[c]{@{}l@{}} 1 - https://docs.datadoghq.com/\\dora\_metrics/?tab=apicurl \end{tabular}  \\
    &    & \cellcolor[HTML]{ECF4FF}manual & \cellcolor[HTML]{FFCCC9}snowballing & \begin{tabular}[c]{@{}l@{}} 2 - https://docs.datadoghq.com/\\tests/ \end{tabular}  \\
    &    & \cellcolor[HTML]{ECF4FF}manual & \cellcolor[HTML]{FFCCC9}snowballing & \begin{tabular}[c]{@{}l@{}} 3 - https://www.datadoghq.com/\\blog\\/best-practices\\-for-ci-cd-monitoring/ \end{tabular}  \\
    &    & \cellcolor[HTML]{ECF4FF}manual & \cellcolor[HTML]{FFCCC9}snowballing & \begin{tabular}[c]{@{}l@{}} 4 - https://plugins.jenkins.io/\\datadog/ \end{tabular}  \\
    &    & \cellcolor[HTML]{ECF4FF}manual & \cellcolor[HTML]{FFCCC9}snowballing & \begin{tabular}[c]{@{}l@{}} 5 - https://docs.datadoghq.com/\\continuous\_delivery/explorer/ \end{tabular}  \\
    &    & \cellcolor[HTML]{ECF4FF}manual & \cellcolor[HTML]{FFCCC9}snowballing & \begin{tabular}[c]{@{}l@{}} 6 - https://docs.datadoghq.com\\/continuous\_integration/\\explorer/?tab=testruns \end{tabular}  \\
    &    & \cellcolor[HTML]{ECF4FF}manual & \cellcolor[HTML]{FFCCC9}snowballing & \begin{tabular}[c]{@{}l@{}} 7 - https://docs.datadoghq.com/\\monitors/types/ci/?tab=pipelines \end{tabular}  \\
        \hline 

\multirow{2}{*}{2}
& \multirow{2}{*}{MEERCODE}  
        & \cellcolor[HTML]{ECF4FF}manual & \cellcolor[HTML]{FFCCC9}snowballing & \begin{tabular}[c]{@{}l@{}} 8 - https://github.com/\\marketplace/\\meercode-ci-monitoring \end{tabular}  \\
    &    & \cellcolor[HTML]{ECF4FF}manual & \cellcolor[HTML]{FFCCC9}snowballing & \begin{tabular}[c]{@{}l@{}} 9 - https://meercode.io/ \end{tabular}  \\
        \hline 

\multirow{2}{*}{3}
& \multirow{2}{*}{CODE CLIMANTE}  
        & \cellcolor[HTML]{ECF4FF}manual & \cellcolor[HTML]{FFCCC9}snowballing & \begin{tabular}[c]{@{}l@{}} 10 - https://codeclimate.com/\\velocity/dora-metrics-in-velocity \end{tabular}  \\
    &    & \cellcolor[HTML]{ECF4FF}manual & \cellcolor[HTML]{FFCCC9}snowballing & \begin{tabular}[c]{@{}l@{}} 11 - https://codeclimate.com/\\blog/\\software-engineering-cycle-time \end{tabular}  \\
        \hline 

\multirow{2}{*}{4}
& \multirow{2}{*}{DEEPSOURCE}  
        & \cellcolor[HTML]{ECF4FF}manual & \cellcolor[HTML]{FFCCC9}snowballing & \begin{tabular}[c]{@{}l@{}} 12 - https://deepsource.io \end{tabular}  \\
    &    & \cellcolor[HTML]{ECF4FF}manual & \cellcolor[HTML]{FFCCC9}snowballing & \begin{tabular}[c]{@{}l@{}} 13 - https://docs.deepsource.com/\\docs/issues\#security-issues \end{tabular}  \\
        \hline  

\multirow{1}{*}{5}
& \multirow{1}{*}{COVERALLS}  
        & \cellcolor[HTML]{ECF4FF}manual & \cellcolor[HTML]{FFCCC9}snowballing & \begin{tabular}[c]{@{}l@{}} 14 - https://docs.coveralls.io/ \end{tabular}  \\
        \hline   

\multirow{1}{*}{6}
& \multirow{1}{*}{COVERITY}  
        & \cellcolor[HTML]{ECF4FF}manual & \cellcolor[HTML]{FFCCC9}snowballing & \begin{tabular}[c]{@{}l@{}} 15 - https://scan.coverity.com/ \end{tabular}  \\
        \hline   

\multirow{2}{*}{7}        
& \multirow{2}{*}{SONAR CLOUD}  
        & \cellcolor[HTML]{ECF4FF}manual & \cellcolor[HTML]{FFCCC9}snowballing & \begin{tabular}[c]{@{}l@{}} 16 - https://docs.sonarsource.com/\\sonarcloud/\\digging-deeper/\\metric-definitions/ \end{tabular}  \\
    &    & \cellcolor[HTML]{ECF4FF}manual & \cellcolor[HTML]{FFCCC9}snowballing & \begin{tabular}[c]{@{}l@{}} 17 - https://circleci.com/\\developer/orbs/orb/\\clicklogiq/sonarqube \end{tabular}  \\
        \hline           

\multirow{1}{*}{8}
& \multirow{1}{*}{SOURCE CLEAR}  
        & \cellcolor[HTML]{ECF4FF}manual & \cellcolor[HTML]{FFCCC9}snowballing & \begin{tabular}[c]{@{}l@{}} 18 - https://app.sourceclear.io/\\login \end{tabular}  \\
        \hline    

\multirow{1}{*}{9}
& \multirow{1}{*}{CCMENU}  
        & \cellcolor[HTML]{ECF4FF}manual & \cellcolor[HTML]{FFCCC9}snowballing & \begin{tabular}[c]{@{}l@{}} 19 - http://ccmenu.org \end{tabular}  \\
        \hline  

\multirow{1}{*}{10}
& \multirow{1}{*}{PLATFORMIO}  
        & \cellcolor[HTML]{ECF4FF}manual & \cellcolor[HTML]{FFCCC9}snowballing & \begin{tabular}[c]{@{}l@{}} 20 - https://docs.platformio.org \end{tabular}  \\
        \hline 


\multirow{1}{*}{11}
& \multirow{1}{*}{MONITOR-PRO}  
        & \cellcolor[HTML]{ECF4FF}manual & \cellcolor[HTML]{FFCCC9}snowballing & \begin{tabular}[c]{@{}l@{}} 21 - https://plugins.jenkins.io/\\monitor-pro/ \end{tabular}  \\
        \hline

\multirow{1}{*}{12}
& \multirow{1}{*}{MONITORING}  
        & \cellcolor[HTML]{ECF4FF}manual & \cellcolor[HTML]{FFCCC9}snowballing & \begin{tabular}[c]{@{}l@{}} 22 - https://plugins.jenkins.io/\\monitoring/ \end{tabular}  \\
        \hline

\multirow{1}{*}{13}
& \multirow{1}{*}{REMOTING-OPENTELEMETRY}  
        & \cellcolor[HTML]{ECF4FF}manual & \cellcolor[HTML]{FFCCC9}snowballing & \begin{tabular}[c]{@{}l@{}} 23 - https://plugins.jenkins.io/\\remoting-opentelemetry/ \end{tabular}  \\
        \hline           

\multirow{1}{*}{14}
& \multirow{1}{*}{DOTCOMMONITOR-LOADVIEW}  
        & \cellcolor[HTML]{ECF4FF}manual & \cellcolor[HTML]{FFCCC9}snowballing & \begin{tabular}[c]{@{}l@{}} 24 - https://plugins.jenkins.io/\\dotcommonitor-loadview/ \end{tabular}  \\
        \hline

\multirow{1}{*}{15}
& \multirow{1}{*}{GITHUB-AUTOSTATUS}  
        & \cellcolor[HTML]{ECF4FF}manual & \cellcolor[HTML]{FFCCC9}snowballing & \begin{tabular}[c]{@{}l@{}} 25 - https://plugins.jenkins.io/\\github-autostatus/ \end{tabular}  \\
        \hline

\multirow{1}{*}{16}
& \multirow{1}{*}{EXTERNAL-MONITOR-JOB}  
        & \cellcolor[HTML]{ECF4FF}manual & \cellcolor[HTML]{FFCCC9}snowballing & \begin{tabular}[c]{@{}l@{}} 26 - https://plugins.jenkins.io/\\external-monitor-job/ \end{tabular}  \\
        \hline

\multirow{1}{*}{17}
& \multirow{1}{*}{PULL-REQUEST-MONITORING}  
        & \cellcolor[HTML]{ECF4FF}manual & \cellcolor[HTML]{FFCCC9}snowballing & \begin{tabular}[c]{@{}l@{}} 27 - https://plugins.jenkins.io/\\pull-request-monitoring \end{tabular}  \\
        \hline        

\multirow{1}{*}{18}
& \multirow{1}{*}{NEWRELIC}  
        & \cellcolor[HTML]{ECF4FF}manual & \cellcolor[HTML]{FFCCC9}snowballing & \begin{tabular}[c]{@{}l@{}} 28 - https://github.com/newrelic/\\nr-jenkins-plugin\#readme \end{tabular}  \\
        \hline

\multirow{1}{*}{19}
& \multirow{1}{*}{METRIC-DATADOG}  
        & \cellcolor[HTML]{ECF4FF}manual & \cellcolor[HTML]{FFCCC9}snowballing & \begin{tabular}[c]{@{}l@{}} 29 - https://github.com/\\jenkins-infra/\\plugins-wiki-docs/blob/\\master/README.md \end{tabular}  \\
        \hline

\end{tabular}
\end{table*}

\begin{table*}
\caption{Official documents analyzed from third-party tools (continuation) }
\label{tab:third-party-oficial-documents-2}
\begin{tabular}{lllll}
\hline
\textbf{\#} & \textbf{third-party tool} & \textbf{Method} & \textbf{Scope} & \textbf{Analyzed Documents}        \\ \hline


\multirow{1}{*}{20}
& \multirow{1}{*}{AMAZON CLOUDWATCH}  
        & \cellcolor[HTML]{ECF4FF}manual & \cellcolor[HTML]{FFCCC9}snowballing & \begin{tabular}[c]{@{}l@{}} 30 - https://aws.amazon.com/pt/\\cloudwatch/features/\\application-monitoring/ \end{tabular}  \\
        \hline

\multirow{1}{*}{21}
& \multirow{1}{*}{GOOGLE CLOUD MONITORING}  
        & \cellcolor[HTML]{ECF4FF}manual & \cellcolor[HTML]{FFCCC9}snowballing & \begin{tabular}[c]{@{}l@{}} 31 - https://cloud.google.com/\\monitoring\#documentation \end{tabular}  \\
        \hline        

\multirow{1}{*}{22}
& \multirow{1}{*}{AZURE MONITOR}  
        & \cellcolor[HTML]{ECF4FF}manual & \cellcolor[HTML]{FFCCC9}snowballing & \begin{tabular}[c]{@{}l@{}} 32 - https://learn.microsoft.com/\\en-us/azure/azure-monitor/ \end{tabular}  \\
        \hline


\multirow{1}{*}{23}
& \multirow{1}{*}{GRAYLOG}  
        & \cellcolor[HTML]{ECF4FF}manual & \cellcolor[HTML]{FFCCC9}snowballing & \begin{tabular}[c]{@{}l@{}} 33 - https://go2docs.graylog.org/\\5-2/home.htm \end{tabular}  \\
        \hline

\multirow{1}{*}{24}
& \multirow{1}{*}{ELK STACK}  
        & \cellcolor[HTML]{ECF4FF}manual & \cellcolor[HTML]{FFCCC9}snowballing & \begin{tabular}[c]{@{}l@{}} 34 - https://www.elastic.co/guide/en\\/elastic-stack/current/index.html \end{tabular}  \\
        \hline        

\multirow{1}{*}{25}
& \multirow{1}{*}{PROMETHEUS}  
        & \cellcolor[HTML]{ECF4FF}manual & \cellcolor[HTML]{FFCCC9}snowballing & \begin{tabular}[c]{@{}l@{}} 35 - https://prometheus.io/\\docs/introduction/overview/ \end{tabular}  \\
        \hline

\multirow{1}{*}{26}
& \multirow{1}{*}{NAGIOS}  
        & \cellcolor[HTML]{ECF4FF}manual & \cellcolor[HTML]{FFCCC9}snowballing & \begin{tabular}[c]{@{}l@{}} 36 - https://www.nagios.org/ \end{tabular}  \\
        \hline        

\multirow{1}{*}{27}
& \multirow{1}{*}{ZABBIX}  
        & \cellcolor[HTML]{ECF4FF}manual & \cellcolor[HTML]{FFCCC9}snowballing & \begin{tabular}[c]{@{}l@{}} 37 - https://www.zabbix.com/\\documentation/current/en/manual \end{tabular}  \\
        \hline

\multirow{1}{*}{28}
& \multirow{1}{*}{HONEYCOMB}  
        & \cellcolor[HTML]{ECF4FF}manual & \cellcolor[HTML]{FFCCC9}snowballing & \begin{tabular}[c]{@{}l@{}} 38 - https://docs.honeycomb.io/\\integrations/ci-cd/buildevents/ \end{tabular}  \\
        \hline

\multirow{1}{*}{29}
& \multirow{1}{*}{SUMO LOGIC}  
        & \cellcolor[HTML]{ECF4FF}manual & \cellcolor[HTML]{FFCCC9}snowballing & \begin{tabular}[c]{@{}l@{}} 39 - https://help.sumologic.com/\\docs/metrics/ \end{tabular}  \\
        \hline

\multirow{1}{*}{30}
& \multirow{1}{*}{LIGHTSTEP}  
        & \cellcolor[HTML]{ECF4FF}manual & \cellcolor[HTML]{FFCCC9}snowballing & \begin{tabular}[c]{@{}l@{}} 40 - https://docs.lightstep.com/ \end{tabular}  \\
        \hline        

\multirow{1}{*}{31}
& \multirow{1}{*}{SPLUNK}  
        & \cellcolor[HTML]{ECF4FF}manual & \cellcolor[HTML]{FFCCC9}snowballing & \begin{tabular}[c]{@{}l@{}} 41 - https://www.splunk.com/\\en\_us/products\\/infrastructure-monitoring.html \end{tabular}  \\
        \hline

\end{tabular}
\end{table*}

After sampling the texts, we store them in our database and conducted a Thematic Analysis~\cite{maguire2017doingThematicAnalysis}, where the aforementioned CI practices guided our process of generating themes. To identify CI practices, we examine the CI metrics monitored by these services.

The document analysis and thematic analysis were conducted to identify citations to CI metrics monitoring across the examined documents. The process involved open coding, performed manually by the first author, through a thorough examination of the citations of CI metrics. The core steps began with an initial open coding phase where the documents were read in detail, and codes were generated to represent the monitored metrics. We identify 152 citations (codes) of CI metrics related to monitoring practices across the 70 analyzed documents of CI services (as shown in Table \ref{tab:ci-servers-oficial-documents} and Table \ref{tab:ci-servers-oficial-documents-continuation} ) and 43 citations (codes) of CI metrics related to monitoring practices across the 41 analyzed documents of third-party tools (as shown in Table \ref{tab:third-party-oficial-documents} and Table \ref{tab:third-party-oficial-documents-2}). After, a standardization process was applied to the identified CI metrics citations. This process involved the follow steps: 

\begin{enumerate}
\item We convert all identified citations to lowercase.

\item We eliminate variations in metric nomenclature across different CI services. For example, the citation ``mean time to resolution'' found in a \textsc{GHActions} document was standardized to ``mean time to \textbf{recovery}'', which is the more prevalent term\footnote{https://www.leanix.net/en/wiki/vsm/dora-metrics}. Similarly, the citation ``Lead time to change'' was adjusted to ``lead time \textbf{for} change''.

\item We also employ CI metrics definitions, when available, to facilitate standardization. For instance, the \textsc{Cicle CI} documentation\footnote{https://circleci.com/blog/how-to-measure-devops-success-4-key-metrics/} defines ``Success Rate'' as \textit{``the number of passing runs divided by the total number of runs over a period of time.''} This definition closely aligns with the CI practice used in our survey: ``Build Health'', hence we convert all citation of ``Success Rate'' to  ``Build Health''. Table \ref{tab:standardization_monitoring_references} provides an overview of all standardization procedures conducted. 

\item Subsequent to this process, we eliminate redundant CI metrics citations.

\end{enumerate}

These standardized metrics citations culminated in a final total of 32 unique themes, they were then grouped into high-level themes, with seven themes associated with CI services and nine themes linked to third-party tools, reflecting the tools where the metrics were observed.

\begin{table*}
\caption{Standardization of CI metrics Citations}
\label{tab:standardization_monitoring_references}
\begin{tabular}{llll}
\hline
\textbf{\#} & \textbf{CI metric citations} & & \textbf{standardized themes}        \\ \hline

1 & Success rate of the main branch       & $\rightarrow$     &   build health \\ \hline
2 & Success rate                          & $\rightarrow$     &   build health \\ \hline
3 & Failure rate                          & $\rightarrow$     &   build health \\ \hline
3 & success/failure rate                  & $\rightarrow$     &   build health \\ \hline
4 & Workflow Duration                     & $\rightarrow$     &   build duration  \\ \hline
5 & mean time to resolution               & $\rightarrow$     &   mean time to recovery \\ \hline
6 & Lead time to change                   & $\rightarrow$     &   lead time for changes \\ \hline
7 & Time to restore service               & $\rightarrow$     &   mean time to recovery \\ \hline
8 & Deploys                               & $\rightarrow$     &   number of deploys \\ \hline
9 & number of deployments                 & $\rightarrow$     &   number of deploys \\ \hline
10 & Time to deploy an application        & $\rightarrow$     &   deploy duration \\ \hline
11 & Unit tests duration                  & $\rightarrow$     &   test duration \\ \hline
12 & Time to test an application          & $\rightarrow$     &   test duration \\ \hline
13 & Unit test failures                   & $\rightarrow$     &   test health \\ \hline
14 & Unit test success density            & $\rightarrow$     &   test health \\ \hline
15 & test success/failure rate            & $\rightarrow$     &   test health \\ \hline
16 & retention                            & $\rightarrow$     &   team retention \\ \hline
17 & anti-patterns                        & $\rightarrow$     &   code quality \\ \hline
18 & code defects                         & $\rightarrow$     &   code quality \\ \hline
19 & build cycle time                     & $\rightarrow$     &   cycle time \\ \hline
20 & build Lead Time                      & $\rightarrow$     &   lead time \\ \hline
21 & build pass rate                      & $\rightarrow$     &   build health \\ \hline
22 & Lead time (CICLE CI)                 & $\rightarrow$     &   lead time for changes \\ \hline
23 & Change fail percentage               & $\rightarrow$     &   change failure rate \\ \hline
24 & throughput                           & $\rightarrow$     &   build activity \\ \hline
25 & Duration                             & $\rightarrow$     &   build duration \\ \hline
26 & median build time                    & $\rightarrow$     &   build duration \\ \hline
27 & last build time                      & $\rightarrow$     &   build duration \\ \hline
28 & Time to Recovery                     & $\rightarrow$     &   mean time to recovery \\ \hline
29 & Critical Vulnerabilites Over time	  & $\rightarrow$     &   security vulnerabilities \\ \hline
30 & High Vulnerabilites Over time        & $\rightarrow$     &   security vulnerabilities \\ \hline
31 & Vulnerabilities                      & $\rightarrow$     &   security vulnerabilities \\ \hline
32 & Security Test reports                & $\rightarrow$     &   security vulnerabilities \\ \hline
33 & Security checks                      & $\rightarrow$     &   security vulnerabilities \\ \hline

\end{tabular}
\end{table*}

Similarly to RQ1, during the analysis, we encountered metrics beyond the scope of CI. Therefore, we classified these practices more generally as ``DevOps metrics''. Among them, the DORA metrics were the most frequently mentioned in the official documentation of CI services. The definition of these metrics are:

\begin{itemize}
\item {\textbf{Deployment Frequency (DF)}: This metric refers to the frequency of successful releases to production, measuring how often a company deploys code successfully. How quickly do we release new versions of the system successfully?} 
\item {\textbf{Lead time for changes (LT)}: This metric measures the time that passes for committed code to reach production. How fast does a committed code reach production?}
\item {\textbf{Mean time to recovery (MTTR)}:  This metric measures the time it takes for a service to recovery from a failure. It encompasses the time between the failure being reported until your fix is released into production. How quickly can we fix a bug in production?} 
\item {\textbf{Change failure rate (CFR)}: This metric captures the percentage of changes that were made to a code that resulted in any kind of production failure. How frequently do we throw errors to production?}
\end{itemize}

Finally, as well as in RQ1, the extracted codes and themes were reviewed by another author. We also calculated the agreement score between the two sets of labels using the Cohen Kappa statistical coefficient~\cite{cohen1968weighted}, which measures the agreement between two ``judges''.  We obtained a score of 0.45, indicating ``Moderate Agreement''. Due to the low value of the Cohen's Kappa coefficient for the Document Analysis of RQ3, a third author reviewed the disagreements between the first author and the second author, with the final result being the code obtained from this second round of review. After the second round of review, we improved the Cohen Kappa statistical coefficient to 0.63, which indicates ``Substantial agreement''.


\section{Results}  \label{Results} 

In this section, we present our results for each research question.

\subsection*{\textbf{RQ1: Do developers express the need to monitor CI practices in PRs?}}

\begin{figure}[!htbp]
  \centering
  \includegraphics[width=\linewidth]{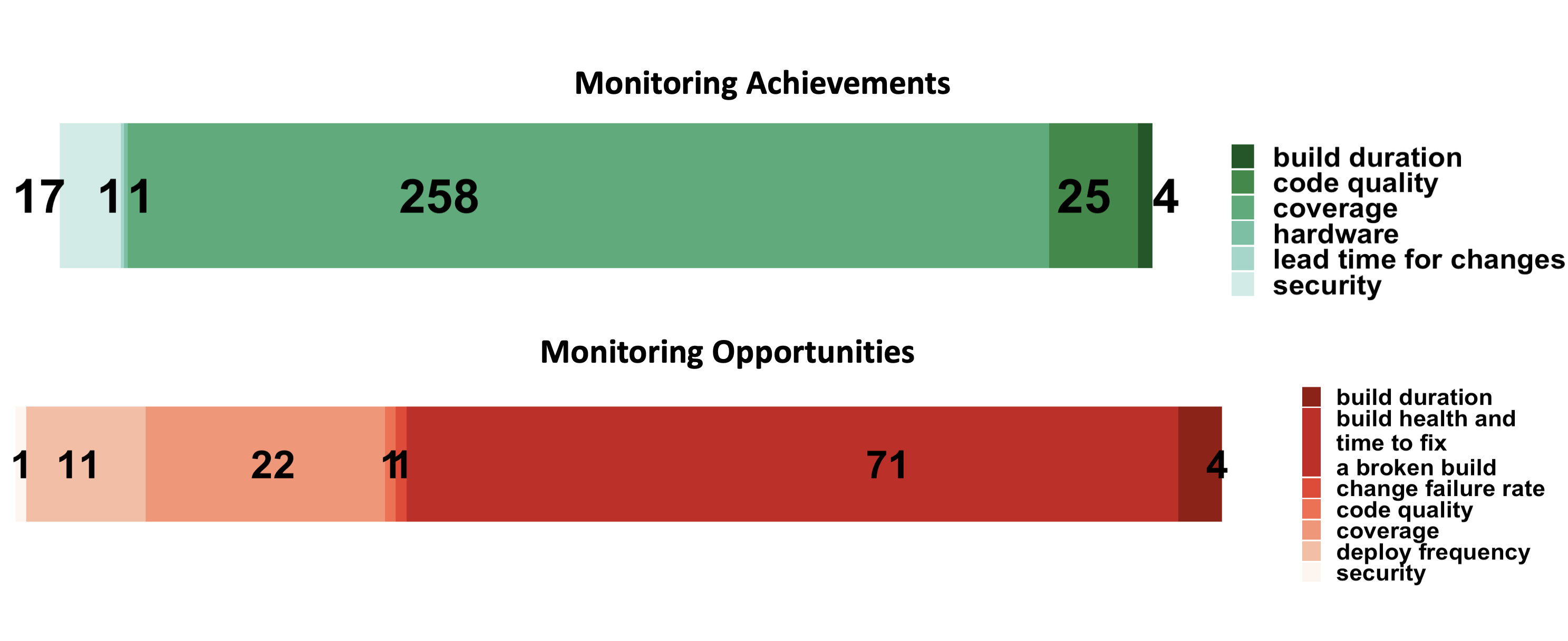}
  \caption{Themes that emerged from PR comments}
  \label{fig:themes_pr_comments}
\end{figure}

The themes generated for our Thematic Analysis, which highlight key patterns and insights derived from the data, are presented in Figure~\ref{fig:themes_pr_comments}.

During the analysis, we find some metrics outside the scope of CI, and we thus classify such practices under the more general ``DevOps metrics'' umbrella.

We observe that developers are mainly monitoring 1) Coverage (e.g., \textit{``... will increase coverage by 0.02''}), 2) Code quality (e.g., \textit{``SonarCloud Quality Gate failed... new-duplicated
-lines-density''}), and 3) security practices (e.g., \textit{``new alerts: 1 for Useless assignment to local variable''}). Coverage was the most frequently monitored  metrics, with over 80\% of themes ($\frac{258}{306}$). \textsc{CodeCov} was the main tool used in most projects, responsible for the majority of coverage monitoring instances ($\frac{206}{258}$).

As shown in Figure \ref{fig:themes_pr_comments}, the main opportunities are for monitoring ``build health'' and the ``time to fix a broken build.'' The analyzed comments indicate that developers face build failures without a sense of the current build health. For example, Comment 2658 states: \textit{``You need to rebase on top of master to fix CI failures.''} Monitoring ``Build Health'' help developers generate fewer errors in private builds, while monitoring the ``time to fix a broken build'' help developers prioritize the fix of builds and improve release quality.

Our results suggest that there exists a concern with monitoring CI practices, although these constitute a small minority $\frac{432}{4892}$ ( or 8\%) of discussions held during the development of the analyzed projects. Whether this is an indication of lack of concern regarding monitoring cannot be answered by our methodological approach applied here in RQ1, since natural discussions through online comments cannot be used to measure the {\bfseries exact level} of concern developers would have regarding a specific issue (e.g., monitoring a CI practice). As such, we contact developers in RQ2 to seek direct opinions regarding the importance of monitoring CI practices.

\subsection*{\textbf{RQ2: What is the perceived importance of monitoring CI practices?}}
\label{RQ2_result}

The results of the survey questions are presented in this section.

\subsection*{Demographics}

\begin{figure}[!htbp]
  \centering
  \includegraphics[width=\linewidth]{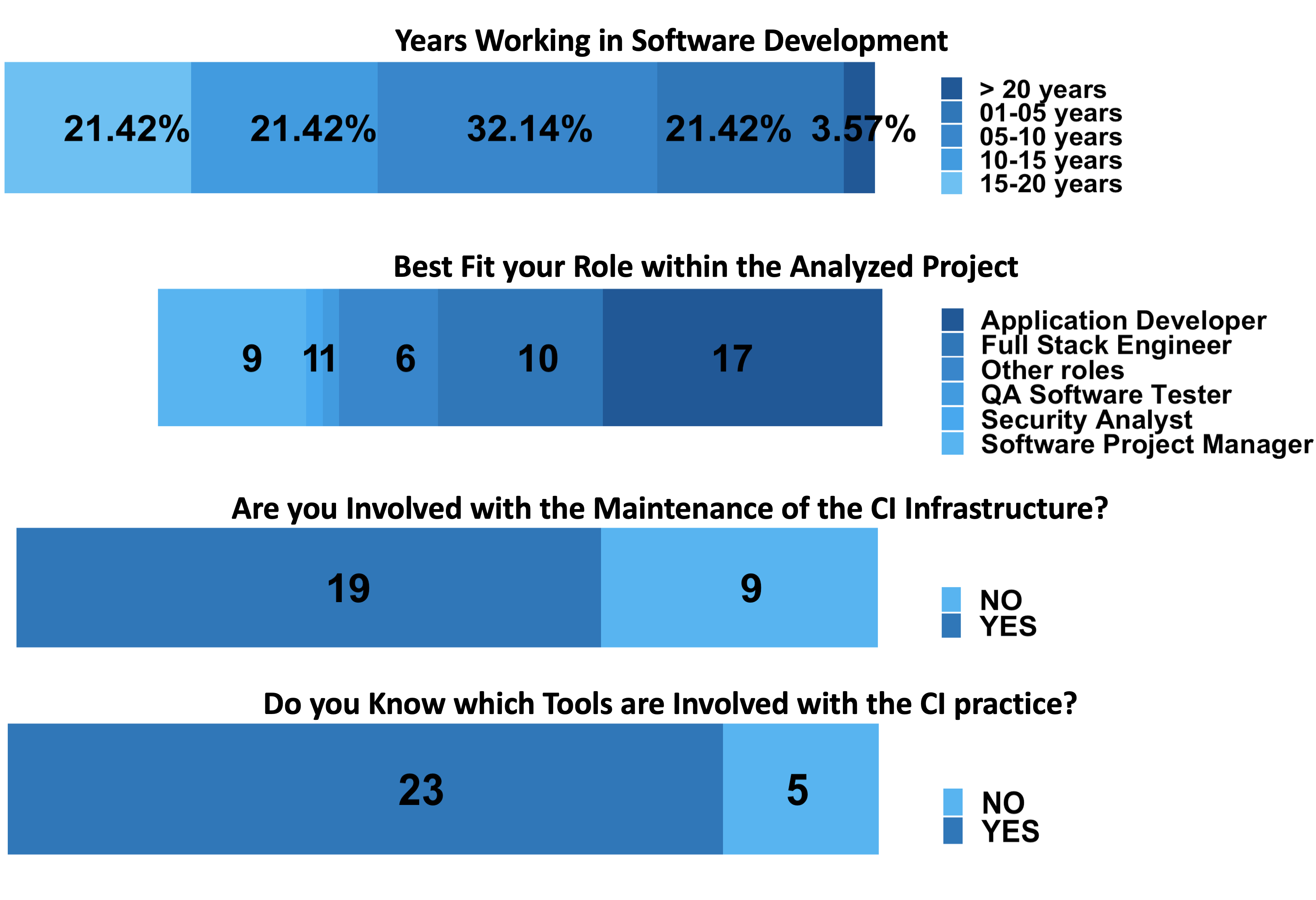}
  \caption{Demographic data of the survey}
  \label{fig:demographics_all}
\end{figure}

Figure \ref{fig:demographics_all} shows the demographic data of the survey. As it suggests, among the subjects, $\frac{13}{28}$ (or 46\%) of software developers have more than 10 years of experience, $\frac{9}{28}$ (or 32\%) have 5 to 10 years of experience, and $\frac{6}{28}$ (or 22\%) have 1 to 5 years of experience working in software development. $\frac{17}{28}$ ( or 60\%) are application developers, $\frac{10}{28}$ ( or 35\% ) are full-stack engineers, $\frac{9}{28}$ (or 32\%) are software project managers, and $\frac{6}{28}$ (or 21\%) have other roles, while only $\frac{2}{28}$ (or 7\%) are QA or security analysts. Most of the developers who answered the survey $\frac{19}{28}$ ( or 67\%) are involved in maintaining the CI infrastructure of the analyzed project. Finally, the majority of developers $\frac{23}{28}$ ( or 82\%) said they know which tools are involved in the CI practice of the analyzed projects. This background demonstrates that we have selected qualified subjects to answer our survey.

\subsection*{Perceptions of CI maturity level}

\begin{figure}[!htbp]
  \centering
  \includegraphics[width=\linewidth]{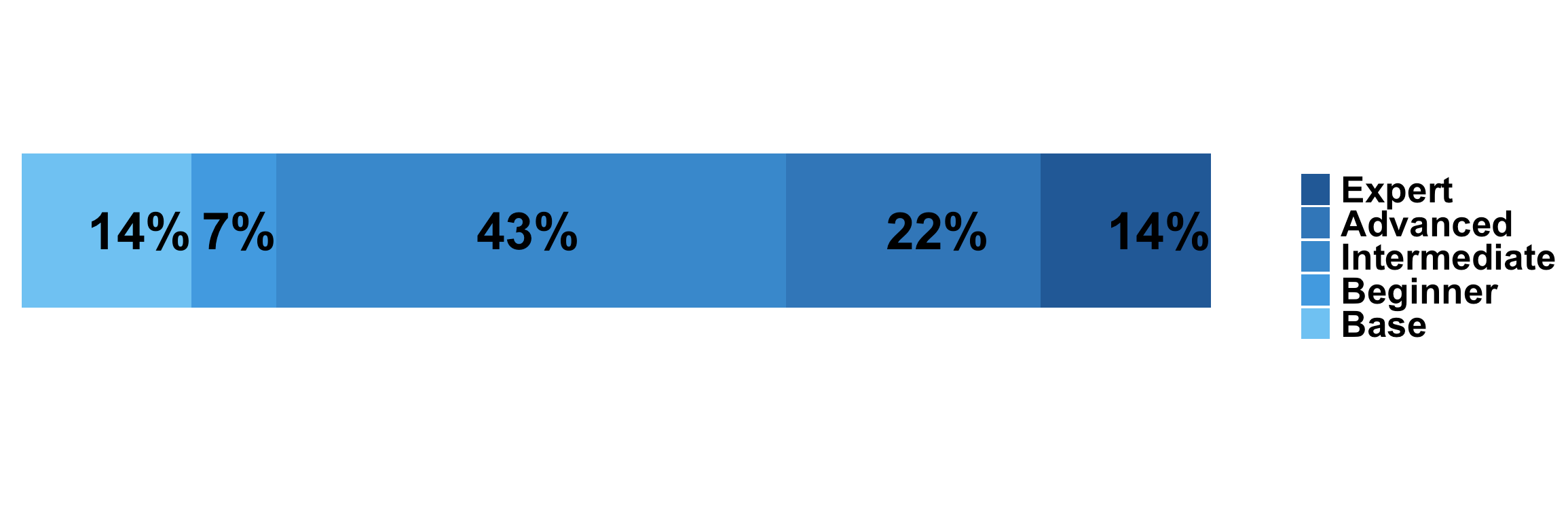}
  \caption{Perceptions of CI maturity level}
  \label{fig:level_ci}
\end{figure}

As shown in Figure \ref{fig:level_ci}, 43\% ($\frac{12}{28}$) of participants consider their project to be in the Intermediate level of CI, whereas 36\% ($\frac{10}{28}$) participants consider their project to be in the Advanced or Expert levels. Lastly, 21\% ($\frac{6}{28}$) of participants consider their projects to be in the Beginner or Base levels.

The answers showed that for most survey participants $\frac{12}{19}$, they consider mainly the automatic execution of the CI build (CI service) and the automatic execution of tests to define the CI level of the project. Among the many reasons used to define the level of CI of the project, Developer 13 said: \textit{``CI builds and run unit tests and BDD tests on every commit.''} Developer 15: \textit{``The test coverage and build/test duration time is the most important metrics to me.''}, Developer 23: \textit{``For each pull request in XXXXX, developer tests and code coverage reports will be automatically triggered..''} Table \ref{tab:creteria-ci-level} summarizes criteria cited by developers.

\begin{table*}[!htbp]
\caption{Criteria used to define project CI level}
\label{tab:creteria-ci-level}
\begin{tabular}{lll}
\hline
\textbf{\#} &      \textbf{Createria}                     &   \textbf{Mentions}                \\ \hline
 1 & Automatic testing (unit/integration/e2e)             &   \multicolumn{1}{r}{10}           \\ \hline 
 2 & Automatic builds                                     &   \multicolumn{1}{r}{8}            \\ \hline
 3 & Manual deployment process                            &   \multicolumn{1}{r}{3}            \\ \hline
 4 & Automatic docker images builds                       &   \multicolumn{1}{r}{2}            \\ \hline
 5 & Looking at what happens when you make a PR           &   \multicolumn{1}{r}{1}            \\ \hline
 6 & Continuous benchmarking                              &   \multicolumn{1}{r}{1}            \\ \hline
 7 & Automation of release notes compilation              &   \multicolumn{1}{r}{1}            \\ \hline
 8 & Result of any successful CI                          &   \multicolumn{1}{r}{1}            \\ \hline
 9 & CI reports                                           &   \multicolumn{1}{r}{1}            \\ \hline
\end{tabular}
\end{table*}


There is no perceived standard of practices or data used to define the maturity level of CI for a project. Our responses show that most developers $\frac{6}{9}$ consider automatic builds and tests execution sufficient to deem their project as having a high maturity level of CI (Advanced or Expert). As such, our responses tend to overlook other CI practices, such as performing frequent commits and maintaining healthy builds, when gauging the maturity level of CI in their projects.


\subsection*{Monitoring CI practices} \label{RQ2}

\begin{figure}[!htbp]
  \centering
  \includegraphics[width=\linewidth]{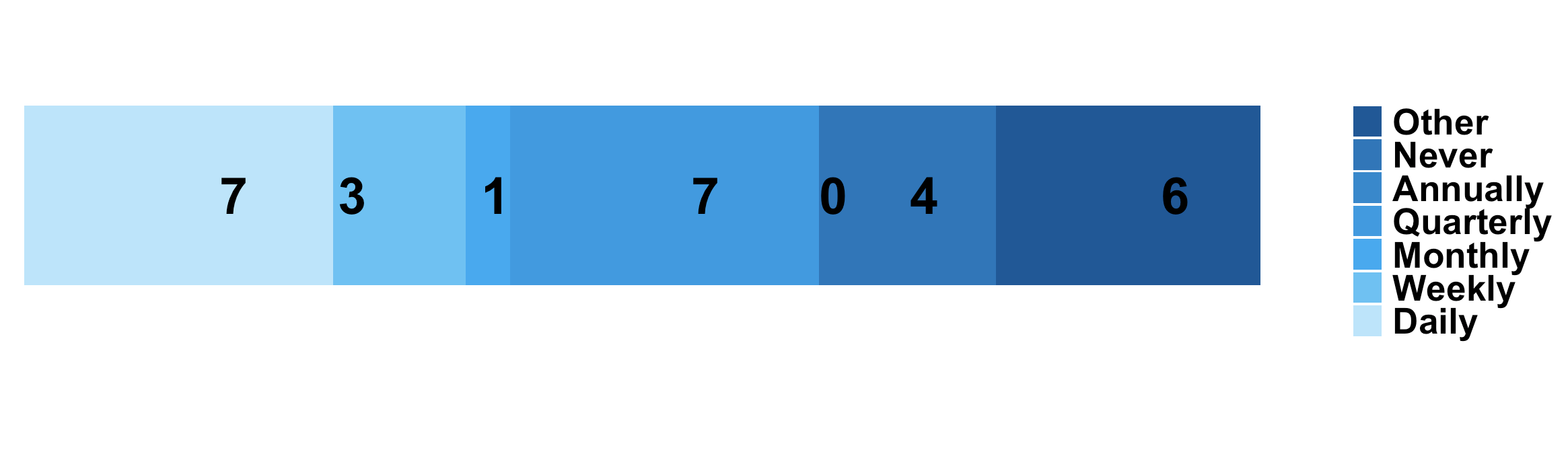}
  \caption{Frequency of CI monitoring}
  \label{fig:ci_frequency_monitoring}
\end{figure}

While participants mostly mention build and test criteria to define their CI maturity level, when asked which practices their projects monitor, respondents mentioned a variety of practices/metrics, as shown in Table \ref{tab:monitored-practices}. When asked to provide reference values for the metrics corresponding to the monitored CI practices in their projects (the most common values for these metrics), only four metrics were mentioned: code coverage, build duration, time to release a new version and commit frequency. Table \ref{tab:monitored-practices-values} summarizes the mentioned values. In relation to frequency, Figure \ref{fig:ci_frequency_monitoring} shows that half of the participants $\frac{14}{28}$ state that monitoring is done daily or quarterly (every 3 months).

\begin{table*}[!htbp]
\caption{Monitored practices/metrics mentioned by the developers}
\label{tab:monitored-practices}
\begin{tabular}{lll}
\hline
\textbf{\#} &      \textbf{CI practice}                      &   \textbf{Mentions}                \\ \hline
 1 & Code coverage                                               &   \multicolumn{1}{r}{14}           \\ \hline
 2 & All tests and inspections must pass                         &   \multicolumn{1}{r}{9}            \\ \hline
 3 & Write automated developer tests                             &   \multicolumn{1}{r}{8}            \\ \hline
 4 & Build duration                                              &   \multicolumn{1}{r}{8}            \\ \hline
 5 & Commit code frequently                                      &   \multicolumn{1}{r}{7}            \\ \hline
 6 & Build health                                                &   \multicolumn{1}{r}{7}            \\ \hline
 7 & Fix broken builds immediately                               &   \multicolumn{1}{r}{7}            \\ \hline
 8 & Don't commit broken code                                    &   \multicolumn{1}{r}{6}            \\ \hline
 9 & Avoid getting broken code                                   &   \multicolumn{1}{r}{6}            \\ \hline
10 & Run private builds                                          &   \multicolumn{1}{r}{4}            \\ \hline
11 & Time of release of new version                              &   \multicolumn{1}{r}{4}            \\ \hline
12 & System/e2e tests                                            &   \multicolumn{1}{r}{1}            \\ \hline
13 & Software quality through the Sonarcloud suite               &   \multicolumn{1}{r}{1}            \\ \hline
14 & Time to fix broken builds                                   &   \multicolumn{1}{r}{1}            \\ \hline
15 & CVEs (Common Vulnerabilities and Exposures) for security    &   \multicolumn{1}{r}{1}            \\ \hline
16 & Vulnerability scan of the resulting docker image            &   \multicolumn{1}{r}{1}            \\ \hline
\end{tabular}
\end{table*}

\begin{table*}[!htbp]
\caption{Reference values to the CI metrics mentioned by developers}
\label{tab:monitored-practices-values}
\begin{tabular}{ll}
\hline
\textbf{CI metric}            & \textbf{reference values}                       \\ \hline

code coverage              & 
\begin{tabular}[c]{@{}l@{}}  \textit{``~60\% coverage''}, 
\\ \textit{``Code coverage 80\%''}, 
\\ \textit{``coverage: >90\%''}, 
\\ \textit{``85\% patch and project coverage ''},
\\ \textit{``coverage: 65\%''}, 
\\ \textit{``coverage: 66\%''}, 
\\ \textit{``code coverage: prefer 90\%, at least 70\%''},  
\\ \textit{``Code coverage: currently 63\%''}, 
\\ \textit{``code coverage - 97\%''}, 
\\ \textit{``Coverage 99\%''} 
\end{tabular}                                                                          \\ \hline

build duration            & 
\begin{tabular}[c]{@{}l@{}} \textit{``roughly a 20 min CI/build cycle''}, 
\\ \textit{``build duration daily''}, 
\\ \textit{``build duration - full CI less than 1 hour''}, 
\\ \textit{``build times less 10 minutes''} 
\end{tabular}                                                                         \\ \hline

time to release a new version       & 
\begin{tabular}[c]{@{}l@{}}  \textit{``time to release a new version 1 month''}, 
\\ \textit{``time to release new version: ~6 months''}, 
\\ \textit{``time to release a new version: 1 weeks''} 
\end{tabular}                                                                        \\ \hline

commit frequency     & 
\begin{tabular}[c]{@{}l@{}}  \textit{``commit frequency - 3 commits in a day''} 
\end{tabular}                                                                         \\ \hline
  
\end{tabular}
\end{table*}

Our responses suggest that some CI metrics are not perceived as important to some participants. For example, Developer 1 expressed: \textit{``All of the above-mentioned items except `Commit code frequently.' I prefer meaningful commits, not frequent ones.''} Developer 25 stated: \textit{``All except 1 (why would commit frequency matter?).''} Additionally, some participants do not see the value of monitoring in their projects. Developer 5 mentioned: \textit{``Most of these metrics are logged in GitHub, although we usually don't analyze the trends or discuss the practices too much as a team, since many of the tracked values fluctuate (Avrae is a kind of monolith which leads to things like coverage being difficult to increase).''}



We find evidence that projects with a lower level of CI (``Base'' and ``Beginner''), monitor less CI metrics. When checking which CI metrics these projects monitor, a maximum of 3 metrics were mentioned. Developer 4: \textit{``We don't actively monitor any of these.''}, Developer 17: \textit{``Mainly 5.''} and Developer 3: \textit{``build health.''} Conversely, in projects deemed as more mature (``Expert'' and ``Advanced''), there was a tendency to monitor more metrics:  Developer 15: \textit{``code coverage, commit frequency, time of release of a new version, build duration, CVEs for security''}, Developer 9: \textit{``1 (Commit code frequently), 3 (Fix broken builds immediately), 4 (Write automated developer tests), 5 (All tests and inspections must pass ), 7 (All tests and inspections must pass )''}, Developer 12: \textit{``All of the above, except commit frequency and to some extent time of release of a new version (this rather depends on completion of the required new features for the version).''}

\begin{figure}[!htbp]
  \centering
  \includegraphics[width=\linewidth]{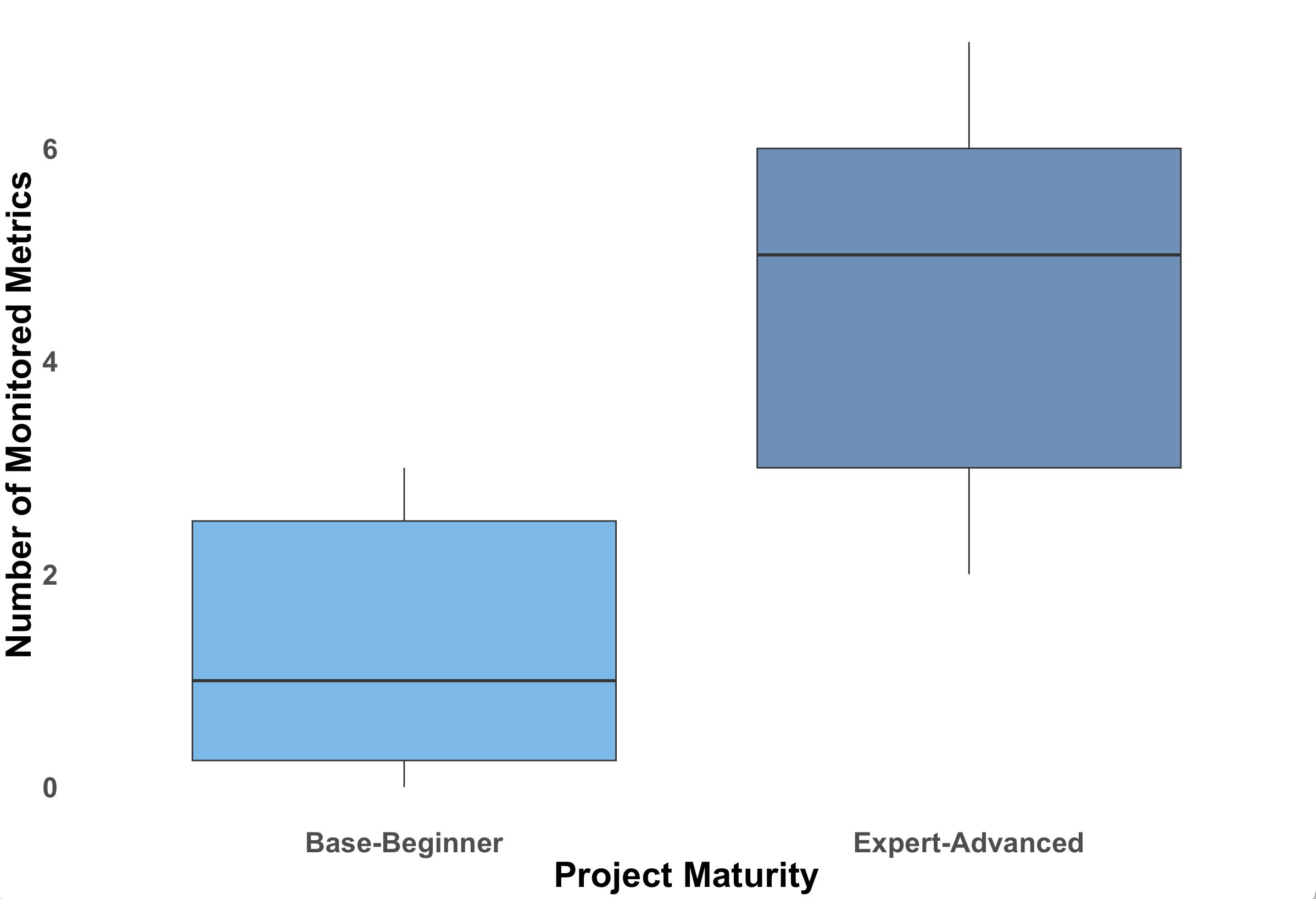}
  \caption{Number of monitored metrics cited by developers x Project Maturity}
  \label{fig:monitoring_distribution}
\end{figure}

Figure \ref{fig:monitoring_distribution} shows the distribution of the number of  metrics mentioned by the developers. The data was organized into two distinct groups: (i) projects classified as ``Expert and Advanced'' and (ii) projects classified as ``Base and Beginner'', excluding ``Intermediate'' projects. The intention of excluding intermediate projects was to specifically highlight the extreme groups where the effects of CI adoption can be more strongly perceived. As observed, developers associated with the ``Expert and Advanced'' group, cited a higher number of monitored metrics.

We applied statistical tests to compare the number of metrics between projects of two previous groups. Our sample for both groups are 16 entries. First, we conducted the Shapiro-Wilk test for normality, which indicated that the data for both groups did not significantly deviate from a normal distribution (p = 0.1494 for Expert-Advanced and p = 0.09315 for Base-Beginner). Levene's test for homogeneity of variances further confirmed the variability of the data in each group is similar (p = 0.3575). Given these results, we performed an independent two-sample t-test, which revealed a statistically significant difference in the number of metrics used between the groups (p = 0.0027). The mean number of metrics for the Expert-Advanced group was 4.5, while the Base-Beginner group had a mean of 1.33, indicating that projects with higher CI maturity tend to use significantly more CI metrics.


\subsection*{Currently Used Tools}

\begin{figure}[!htbp]
  \centering
  \includegraphics[width=\linewidth]{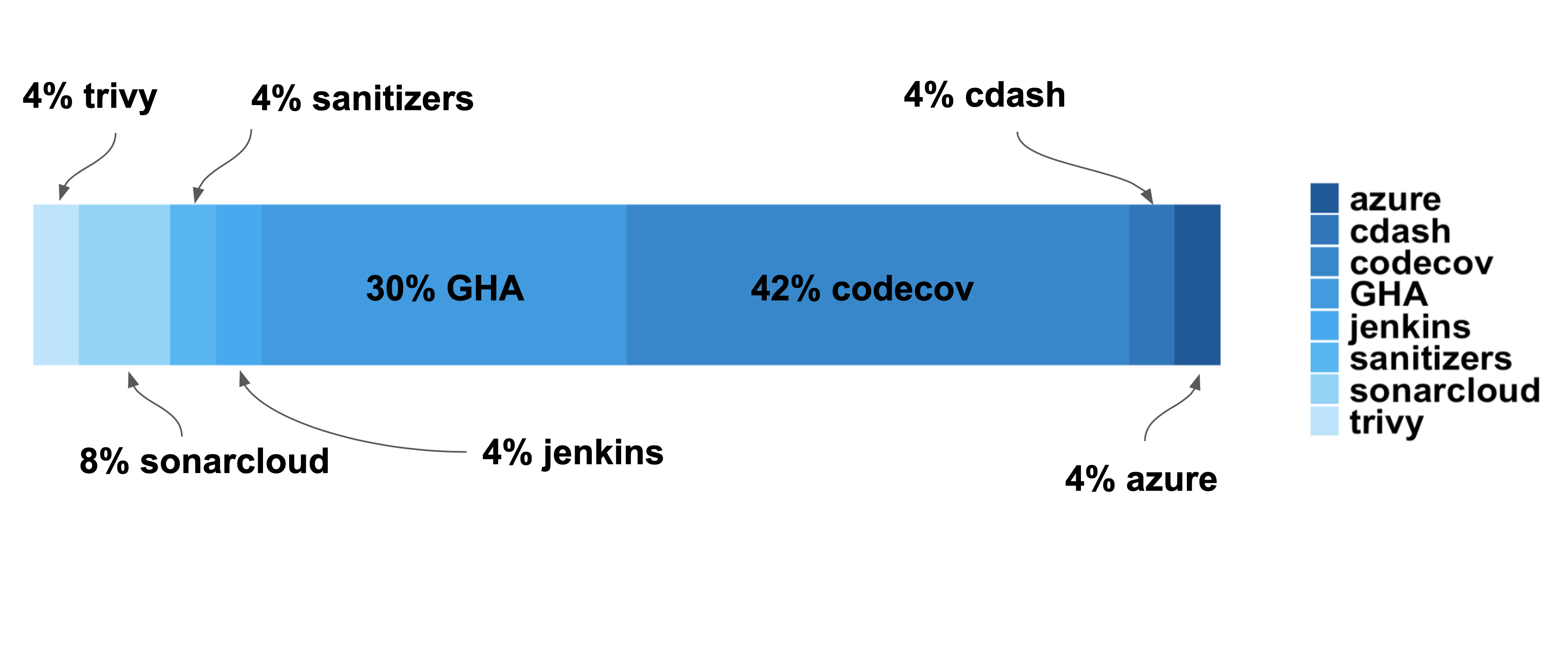}
  \caption{Tools used to Monitor CI practices}
  \label{fig:monitoring_tools}
\end{figure}

Figure~\ref{fig:monitoring_tools} shows an overview of the tools mentioned by participants to perform monitoring tasks in their projects. \textsc{CodeCov} is the most used by 42\% ( or $\frac{11}{26}$ ) of participants. In second place, \textsc{GHActions} itself is used by 30\% (or $\frac{8}{26}$) of participants. \textsc{GHActions} automatically tracks basic build information like ``build time'' and ``failed/passed builds,'' which can be useful for monitoring. \textsc{SonarCloud} comes in third with 8\% ( or $\frac{2}{26}$) of participants using the tool. Based on data from the currently used tools, Developer 20 summarizes the most common monitoring flow for CI practices: \textit{``Codecov for tracking code coverage automatically on each commit. GitHub Actions automatically tracks things like 'build time', 'failed/passed builds', etc.''}, since \textsc{CodeCov} and \textsc{GHActions} are the most used tools.

\begin{sloppypar}
To sum up, most of participants monitor ``code coverage'' using \textsc{CodeCov} $\frac{11}{21}$ (or 52\% ) and information such as ``build duration'' and ``build status'' are monitored using \textsc{GHActions} $\frac{8}{21}$ (or 38\% ). Some developers monitor code quality metrics using \textsc{Sonarcloud} $\frac{2}{21}$ (or 9\% ). Other tools are mentioned once (\textsc{CDash}, \textsc{Trivy}, \textsc{Jenkins}, sanitizers and \textsc{Azure/container-scan}). $\frac{4}{21}$ (or 19\% ) participants mentioned not using any tools to monitor CI metrics. No tools were mentioned to monitor the remaining CI metrics, such as ``time to fix a broken build'', ``build health'' or ``commit activity.'' We noticed that there is a pressing concern regarding security and code quality in the CI pipeline. Developer 8 expressed: \textit{``Azure/container-scan GitHub Action: container best practices and vulnerabilities''}, and Developer 12 said: \textit{``Codecov, SonarCloud, and sanitizers are part of the CI process and ensure that we don't merge untested code to begin with, as part of the review process.''}
\end{sloppypar}

GitHub provides several built-in quality and security tools, such as Dependabot and CodeQL, which are widely used for dependency health monitoring and vulnerability analysis. However, it is notable that these tools did not prominently appear in the survey responses. A plausible explanation is that these tools may have been perceived as part of the GitHub Actions ecosystem, which was mentioned by 30\% of respondents. Since GitHub Actions integrates various CI functionalities, including dependency and security monitoring, respondents may have grouped these tools under the broader GitHub Actions umbrella. Another possibility is that the survey's focus on ``CI metrics'' and it is possible that these tools were perceived as outside the scope of the CI process. This emphasis might have overshadowed the mention of tools like Dependabot and CodeQL in the response.

Our findings corroborate the results from RQ1. In RQ1, we found that $\frac{237}{285}$  (or 83\%) of the monitoring opportunities suggested the monitoring of ``Coverage,'' of which $\frac{205}{237}$ (or 86\%) were associated with the \textsc{CodeCov} tool. \textsc{Sonarcloud} was the third most used tool among survey respondents, which is probably why ``Code Quality'' and ``Security'' practices represent the second and third strongest evidence of monitoring in PR comments, with respectively 25 and 17 mentions. Code quality and security are typically monitored by tools such as \textsc{Sonarcloud}.


\subsection*{Evaluation of CI metrics} \label{RQ3}


When asked about the importance of monitoring the CI metrics presented in our survey (see Figure~\ref{fig:practices_evolution}), most participants $\frac{20}{28}$ (or 71\%), provided favourable responses. Developer 7 stated: \textit{``Yes, these metrics will provide valuable information in the status of our project over time.''}  Developer 15: \textit{``yes, it can make the development more efficient with high quality''},  Developer 24: \textit{``Yes, especially build duration, build health, and time to fix broken builds''}, Developer 20: \textit{``Absolutely, it would help show the health of the project, both for the active maintainers, but also for new people evaluating the project to determine if they want to use it..''} 

Other participants found the metrics useful but not critical for their projects. Developer 9 mentioned: \textit{``In other projects these measurements are important, but in this Open Source project most of the contributors just work on the project when they have some spare time. The information is for sure useful, but not critical for us.''} Some participants found only part of the metrics useful. Developer 25 mentioned: \textit{``depends on the practice, for example `commit frequency' is not important at all, but fixing a broken build is.''} Only 2 developers found the information about CI metrics evolution not useful at all. Developer 26 expressed: \textit{``Not really. I don't think they are important because I know what needs to get done and time is the biggest limit.''}, while Developer 21 said an emphatic \textit{``No.''}

We presume that Developer 21 is a relatively inexperienced developer. He stated that he has 5-10 years of experience with software development but is not involved in the maintenance of the CI infrastructure of the project, and he does not know which tools are involved in the project's CI. He classified his project as having a ``Base'' level of CI maturity. However, because he was not involved in the project's CI configurations and is not familiar with the CI tools used, his answers may not be entirely reliable.

Developer 26 has 10-15 years of experience with software development and classified his project as having an ``Intermediate'' CI maturity level. His project is an embedded JavaScript library designed to include a special kind of graphics in \textsc{html} pages and provide them with extra functionality. Due to the nature of the project (e.g., it is not a high-scale critical project), following all the CI practices may not be deemed necessary, what justifies his answer.



Almost half of respondents $\frac{11}{24}$ (or 46\% ) would like to improve their CI  metrics. Developer 8 stated: \textit{``Absolutely! It saves time in the long run and increases confidence when refactoring is applied.''}, Developer 28: \textit{``Yes, especially build time, fast CI = fast tests and deploy''}, Developer 19: \textit{``Yes. Noticing problems early generally requires less engineering time and creates less distractions compared to noticing quality problems later.''}, Developer 11: \textit{``Build time is something we should definitely worth dedicating time to. Coverage seems okay for us. Comments per change is not that much of an important aspect for us. We require at least two reviewers to approve and usually at our scale, we are positive that the reviewers usually do a good job.''}, Developer 2: \textit{``Yes. Tracking the evolution of practices over time helps a project identify areas for improvement, evaluate changes, ensure consistency, and demonstrate progress to stakeholders.''} and Developer 4: \textit{``Yes, as it affects the quality of our software product.''}


While some participants believe that it is important to invest time in improving the CI metrics presented in our survey, others $\frac{5}{24}$ ( or 21\% ) do not consider it a priority. For example, Developer 22 stated: \textit{``Sure, if someone has time''}, while Developer 9 explained: \textit{``In this open source project, it is not critical because we have people who tend to fix issues promptly. However, in teams creating software for a company, it is much more important to keep the team productive and avoid being hindered by failures in CI.''}


Some participants $\frac{5}{24}$ ( or 21\% ) do not believe that improving these CI  metrics are important. Developer 16 explained: \textit{``For this project, it doesn't make much sense because the current approach seems to be working well''}, while Developer 26 stated: \textit{``I don't see how these metrics are connected to important goals like reducing errors''}



When asked if they find regular updates on the CI metrics useful, 68\% of the developers $\frac{19}{28}$ answered ``Yes''. The responses indicate that the majority of developers consider the studied CI metrics important and would monitor them if provided with the opportunity.



\begin{figure}[!htbp]
  \centering
  \includegraphics[width=\linewidth]{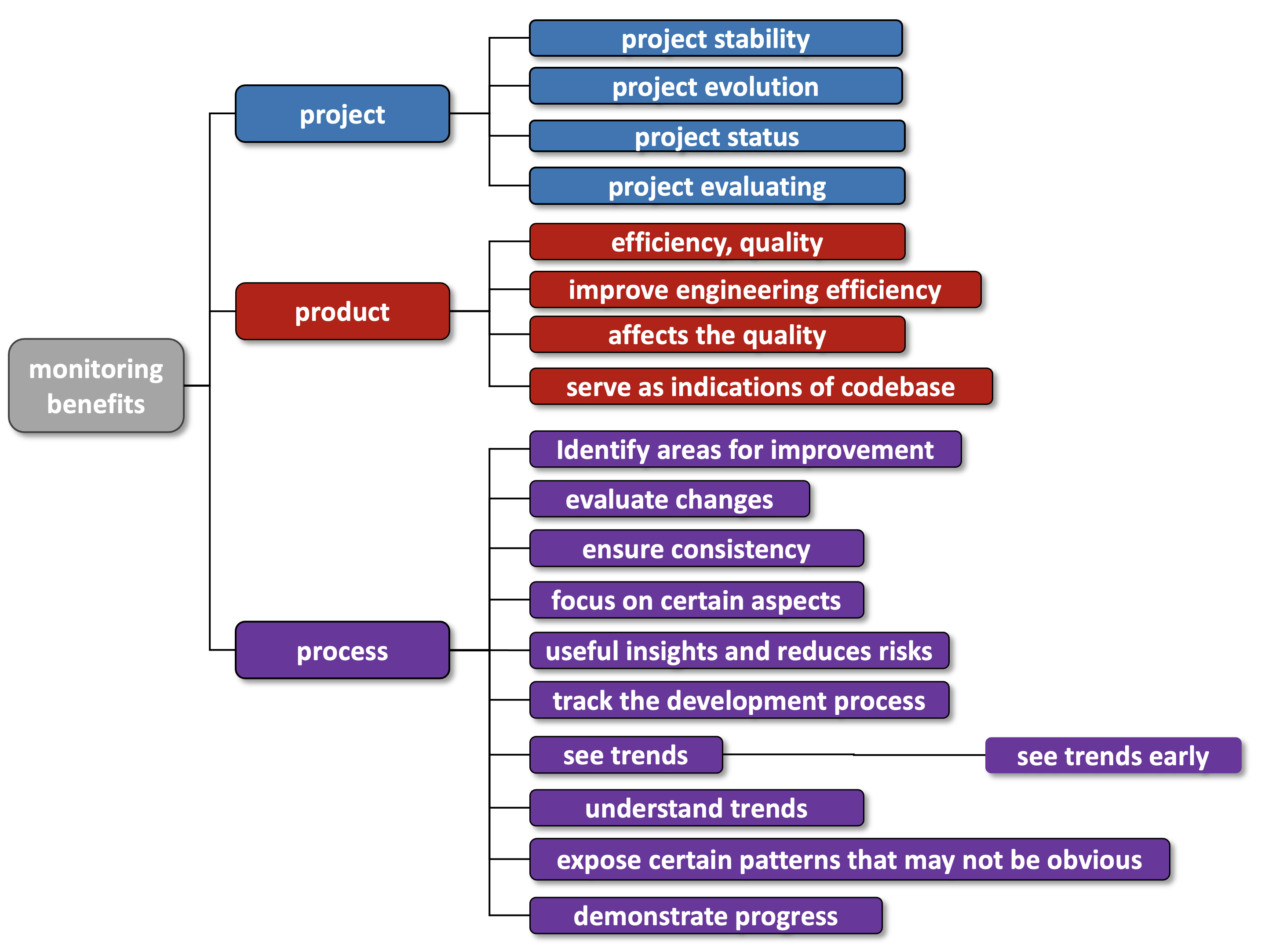}
  \caption{Benefits of monitoring CI metrics}
  \label{fig:themes_monitoring_importance}
\end{figure}

In Figure~\ref{fig:themes_monitoring_importance}, we code the benefits listed by developers when we asked about the importance of monitoring the CI metrics presented in our survey. They were grouped in three main themes: (i) ``benefits for the project'', (ii) ``benefits for the product'' and (iii) ``benefits for  the process''.  We recall that the coding process was simpler and differed from those used in the Document Analysis of RQ1 and RQ3, as it was not reviewed by other authors. The objective was to summarize the benefits listed by developers, in a single and simplified view.

Our findings related to this theme also corroborates our results from RQ1, where we primarily identified opportunities to monitor ``build health'' and ``time to fix a broken build.'' These two metrics were considered important by developers. For example, Developer 24 stated: \textit{``Yes, especially build duration, \textbf{build health, and time to fix broken builds}''}, while Developer 20 mentioned: \textit{``Absolutely. \textbf{It would help show the health of the project}, both for the active maintainers and also for new people evaluating the project to determine if they want to use it.''}


\subsection*{Impact of CI practices}


Most of participants $\frac{22}{24}$ (or 91\% ) agree with the statement that CI impacts project productivity and quality. Developer 8 stated:  \textit{``Absolutely. It helps to increase confidence in the state of the code baseline and functionality, as long as the dev effort to create CD/CI checks is well invested.''} Developer 11 added:  \textit{``Yes, CI is extremely important for us in terms of both code quality and productivity. We need to build and deploy on various hardware architectures and without CI, we would end up spending enormous amounts of time adding fixes for different hardware, and bugs would be impossible to track. Our CI also gives us a quick overview of the issues in our code, helping us quickly fix any problems. In addition to continuous integration, it is important that developer instances can also be quickly spun up to make sure our tests pass. The containers we use for our CI also help us here.''} Developer 7 agreed, stating:  \textit{``Yes, it takes away the boring task of maintaining the project in order to dedicate more time to productive coding.''} Developer 6 also agreed, stating:  \textit{``Yes, mostly when it comes to new developers, CI and automated builds and tests are a good way to avoid breaking the code.''} Only one participant had questioned the benefits of CI when it comes to productivity and quality. Developer 13 stated:  \textit{``Partially? This is hard to say - I think the overall trends look OK, and the current quality and productivity don't seem to have any issues, so it lines up.''} One developer did not know answer the question, Developer 26 stated: \textit{``I don't know.''}


\subsection*{Correlations with Quality Attributes}


Although most developers feel that CI impacts productivity and quality of projects, few developers $\frac{2}{18}$ (or 11\% ) found the statistical correlations between CI metrics and bug-related issues or number of closed PRs useful (see Figure \ref{fig:practice_correlation}). Developer 16: \textit{``Hard to say but it makes sense that more commits lead to more closed pull requests''} and Developer 13: \textit{``Generally one commit = one pull request, so I don't think this particular metric means anything for this project.''} Developer 24: \textit{``No, I did not find these 3 correlations interesting or meaningful. Not sure it will help somehow.''} 

In Developer 24's project, there is a correlation of 0.58 between ``Commits per Weekday'' and ``Bug-related issues.'' We conjecture that a higher number of commits may be producing more bug-related issues, which can potentially indicate lack of testing for the project. Such correlation could generate some sort of alert to be investigated by the project team: \textit{``Why, in my project, does a commit have a significant correlation with generating bug-related issues?.''} Even so, the participant did not find this correlation useful, because they did not understand the implications of the correlations. 

Only 2 developers perceived advantages in the correlations between CI metrics and quality attributes presented in the survey: Developer 28: \textit{``Yes, diagnostic reasons''} and Developer 20: \textit{``Yes, it shows that the CI pipeline is helpful.''}


\subsection*{Considerations about Monitoring}


In this last part of the survey, we asked participants whether there were any CI metrics we missed or whether our participants preferred other types of visualizations for the metrics presented.


$\frac{2}{18}$ (or 11\% ) developers expressed that the metrics and visualizations presented were adequate. Developer 2: \textit{`Line charts can show trends in practice performance over time, while bar charts can compare the performance of different practices. Both are useful for a report on CI practices.''} and Developer 27: \textit{``graphs were perfect.''} $\frac{3}{18}$ (or 16\% ) did not miss any information, they answered a simple: \textit{``No.''} $\frac{2}{18}$ (or 11\% ) developers are not sure which metrics they would like to see monitored. Developer 10: \textit{``Not sure''} and Developer 22: \textit{``I'm not sure.''} $\frac{1}{18}$ (or 5\% ) Developer provided an unhelpful answer and has been discarded: \textit{``CI environments are sometimes really annoying to debug.''} $\frac{10}{18}$ (or 55\% ) suggested improvements to the information contained in the survey. We describe some of the suggestions in the next few paragraphs.

Developer 15, while liked the visualization of the metrics in the survey: \textit{``The pictures above including these 7 charts are interesting to me.''}, suggested including security metrics like image scan, code dependency scan, and configuration scan. Although important, these security metrics are outside the scope of our survey as they do not directly relate to CI but rather to DevOps. Developer 28 suggested including ``deploy time'' as an interesting metrics for CD in future research. 

Participants also discussed other matters related to PR, Issue, and Commit data. Developer 11 stated, \textit{``PR and issue related data seems to be missing. It should be public, so you should be able to gather that from our GitHub project. We use tags to sort them.''} Developer 13 suggested, \textit{``Perhaps commit size over time too? Tracking keywords used in commit messages?.''} Developer 12 also like how the information of seven CI metrics were presented in the survey: \textit{``I think the above information is enough. For our current usage, having these simple graphs available for a daily/weekly tracking of the project status would be sufficient to assign someone to fixing the problem. We do this in a more informal way currently''}, but recommended investigating \textit{``statistics on who reviews and who commits, if there's a change in the balance (less committers or less reviewers) as important. The aim is to ensure knowledge sharing, and that one developer does not 'take over the project' while breaking the information and knowledge sharing that is in place.''} While PR and Issue information were used to compute correlations regarding CI practices in our study, they are not CI metrics and were not the focus of this paper. Both ``Commit size'' and ``Ratio between committers and reviewers,'' as well the consequences of an unbalanced ratio, are interesting aspects to investigate in future research. 

\begin{figure}[!htbp]
  \centering
  \includegraphics[width=\linewidth]{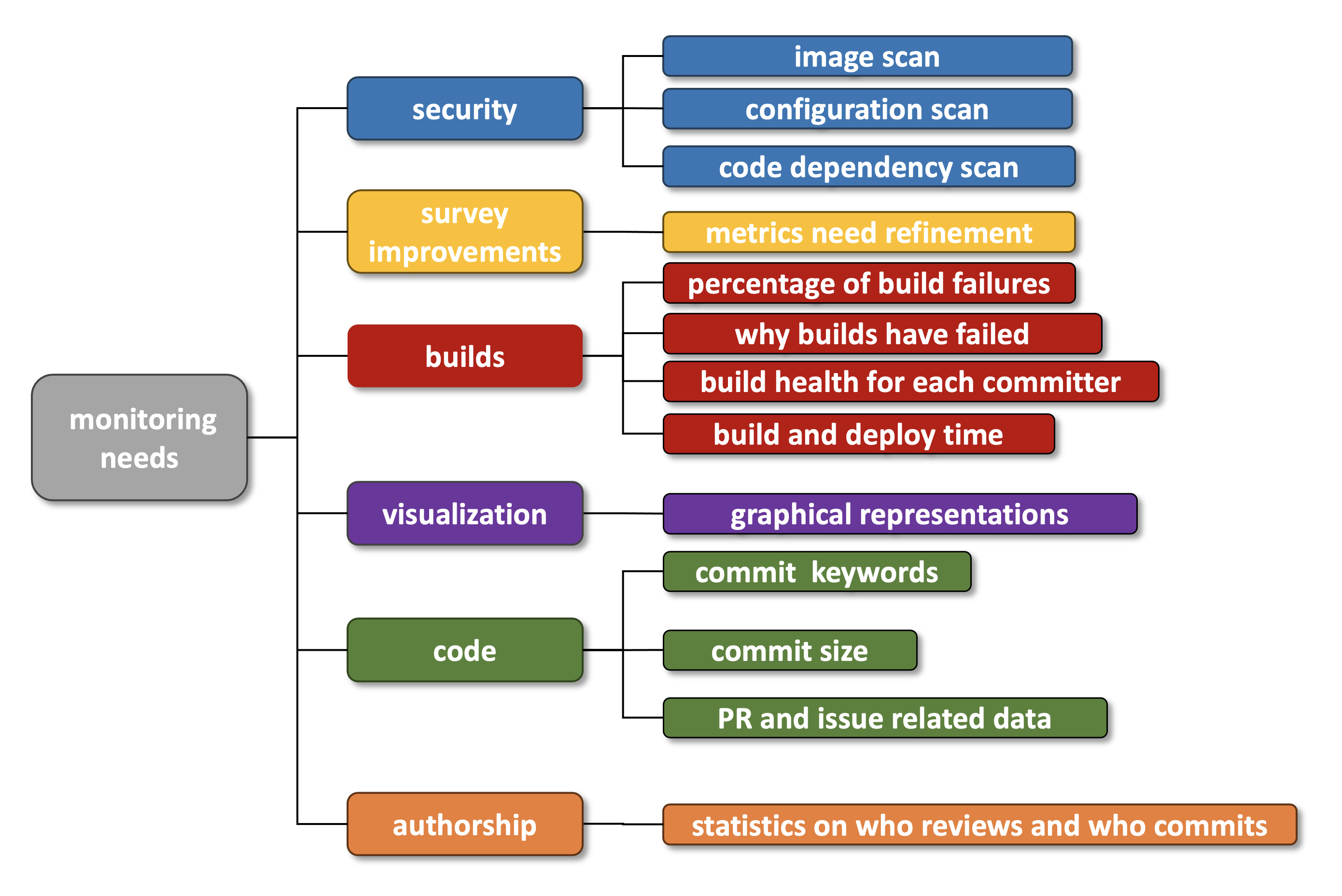}
  \caption{Needs related to monitoring CI metrics}
  \label{fig:themes_monitoring_needs}
\end{figure}

For a broader view of the monitoring needs, in Figure~\ref{fig:themes_monitoring_needs} we show the themes that emerged from the monitoring needs suggested by survey participants. As same way of Figure \ref{fig:themes_monitoring_importance}, we recall that the coding process was simpler and differed from those used in the Document Analysis of RQ1 and RQ3. The objective was to summarize the monitoring needs in a single and simplified view.

\subsection*{\textbf{RQ3: Which monitoring features are currently supported by existing CI tools?}}

\begin{figure}[!htbp]
  \centering
  \includegraphics[width=\linewidth]{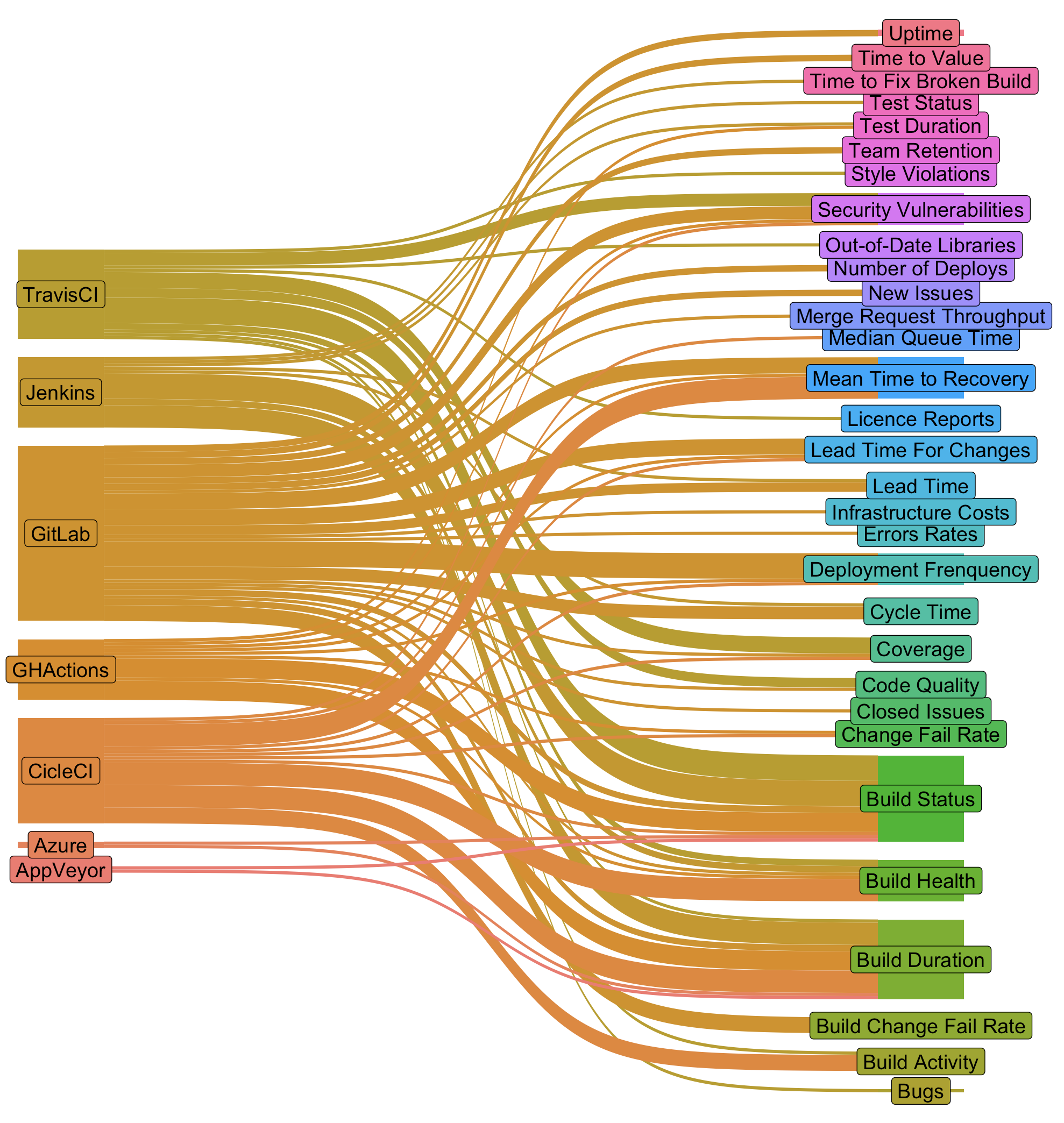}
  \caption{Themes that emerged from CI services documentation}
  \label{fig:themes_ci_servers}
\end{figure}

Figure \ref{fig:themes_ci_servers} illustrates the themes that emerged from the official documentation of the seven most popular CI services \cite{GolzadehCIServer2022}. We emphasize that the results described below are derived from the thematic analysis conducted on the official documentation of the CI services included in this study and are limited to the metrics present in the examined documents.

We observe that \textbf{``Build Duration''} and \textbf{``Build Status''} are the most frequently mentioned themes in the official documentation of CI services. This is due to the fact that all CI services provide information about the time taken for the build and whether the process was successful or not. However, these metrics are fairly rudimentary and provide limited insight into the accurate implementation of Continuous Integration, Continuous Delivery, or DevOps practices. They are insufficient for detecting the proper application of these practices because they do not cover all the essential aspects, such as code quality, build and deploy frequency, and user feedback, among others. \textbf{``Build Health''} emerges as the third most common theme in the documentation of CI services, stemming from the metric ``Build Status''.

In this analysis we find other metrics mentioned outside the scope of CI, they were classified under the umbrella of ``DevOps metrics''. For example, the DORA metrics \cite{DoraMetricsLeanIX2023} have been widely mentioned for the CI services. \textsc{Gitlab}, \textsc{GHActions}, and \textsc{Circle CI} cite these metrics. However, only the \textsc{Gitlab} documentation declare that there is an implementation of a dashboard to monitor these metrics. In May 22, 2023\footnote{https://about.gitlab.com/releases/2023/05/22/gitlab-16-0-released/}, \textsc{Gitlab} has added the ``Value Stream Dashboard'' which included the DORA and Vulnerabilities metrics, but is still accessible only from Premium version\footnote{https://stackoverflow.com/questions/76382478/permission-denied-for-gitlab-dora-metrics-through-api}. \textsc{Gitlab} was the service with the highest number of citations ($\frac{55}{161}$) (or 34\% ) for themes that refer to DevOps metrics monitoring, 1.53 times more than the second place CicleCI with ($\frac{33}{161}$) (or 20\% ). This indicates that Gitlab is the most advanced CI service in terms of supporting the monitoring of DevOps metrics, without the need to use third-party tools.

In the \textsc{GHActions} documentation, we did not find indications of native monitoring tools specifically designed for CI metrics. The existing monitoring tools appear to concentrate on optimizing the CI/CD pipeline by monitoring aspects such as CPU and memory usage within the pipeline. However, \textsc{GHActions} allows the creation of various plugins\footnote{https://github.com/marketplace?type=apps\&category=continuous-integration} that can complement these monitoring tasks.

\textsc{Circle CI} focuses on monitoring four metrics: 1) ``Build Health'', 2) ``Mean Time to Recovery'', 3) ``Build Duration'', and 4) ``Build Activity''. In January 2022, a centralized dashboard\footnote{https://circleci.com/docs/insights/} has been introduced to monitor these four metrics. Although the \textsc{Circle CI} documentation references DORA metrics, there is no clear indication of support for monitoring them. The documentation also cites several third-party tools that can be utilized to monitor DevOps metrics, such as Datadog, Honeycomb, Sumo Logic, Lightstep, New Relic, Splunk, Graylog, ELK stack, and metrics-capturing tools like Prometheus, Nagios, and Zabbix.

\textsc{Jenkins} documentation primarily supports monitoring basic build information, such as ``Build Status'' and ``Build Duration''. With the inclusion of third-party plugins for \textsc{Jenkins}, like DataDog\footnote{https://plugins.jenkins.io/datadog/}, additional metrics are mentioned, including ``Build Health'', ``Build Lead Time'', ``Build Cycle Time'', and ``Time to Fix Broken Build''. This assortment of third-party tools underscores the complexity involved in establishing an environment for monitoring DevOps adoption in the present day.

\textsc{Travis CI} provides support for ``Build status''. We identified several other metrics in the official documentation; however, these metrics are associated with third-party tools integrated into \textsc{Travis CI}. Many of these tools, such as node-build-monitor\footnote{http://marcells.github.io/node-build-monitor/}, are either not functional when accessed or have been deprecated, like ci-dashboard\footnote{https://github.com/ahsayde/ci-dashboard} and team\_dashboard\footnote{http://fdietz.github.io/team\_dashboard/}, which have not been updated for several years.

\textsc{Appveyor} offers a range of integration with various tools, but these integration mainly cover basic build metrics such as ``Build Status'' and ``Build Duration''. \textsc{Azure} monitoring encompasses a complex set of services; however, its focus primarily lies in cloud infrastructure monitoring.

\begin{figure}[!htbp]
  \centering
  \includegraphics[width=\linewidth]{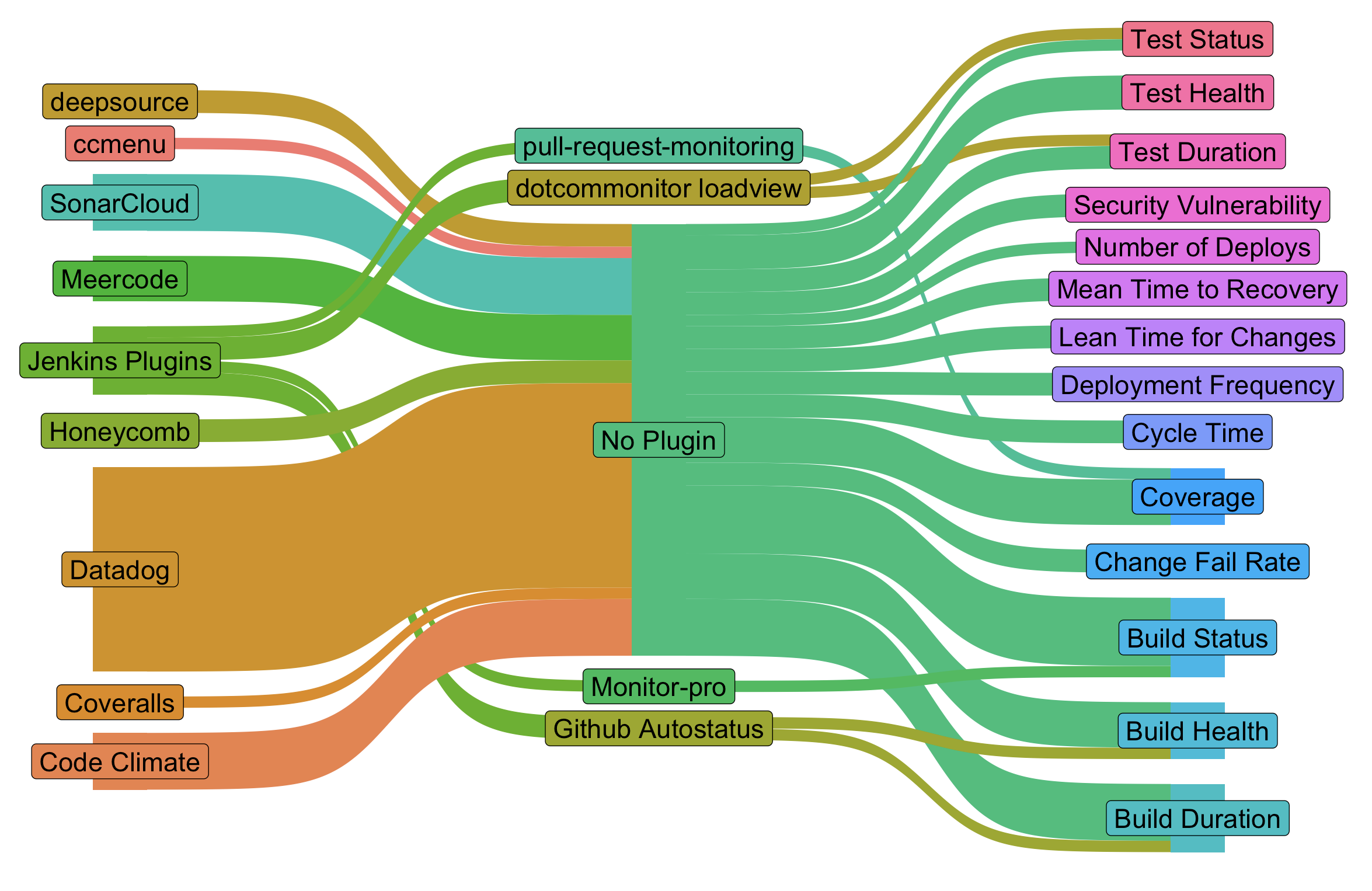}
  \caption{Themes that emerged from third-party tools documentation}
  \label{fig:themes_third_party}
\end{figure}

When examining third-party tools themes, shown in Figure~\ref{fig:themes_third_party}, we observe that, similar to CI services, most of them support monitoring basic pipeline information such as \textbf{``Build Duration''} and \textbf{``Build Status.''} \textsc{DataDog} stands out as the most comprehensive tool for CI monitoring, allowing not only the monitoring of pipeline health but also Coverage, which is related to the quality of the source code generated by the project. Additionally, \textsc{DataDog} supports monitoring DORA metrics, a capability shared with \textsc{Code Climate}. Despite \textsc{DataDog} having very comprehensive set of practices, some practices, such as the project's commit frequency (Commits Per Weekday), initially appear not to be supported by the tool.

Other tools, such as \textsc{SonarCloud}, primarily focus on code quality rather than monitoring the CI pipeline. Cloud tools like \textsc{Google Cloud Monitoring} and \textsc{Azure Monitor} are geared towards monitoring pipeline infrastructure, including CPU and memory usage, but they lack metrics for evaluating the project's CI adoption.

Many monitoring tools, such as \textsc{Graylog}, \textsc{Zabbix}, \textsc{Prometheus}, and the \textsc{ELK stack}, are generic monitoring tools that can be used for CI monitoring but do not implement monitoring of any specific metrics. As the \textsc{ELK stack} documentation describes:  \textit{`` ELK stack obtain data reliably and securely from any source, in any format; then perform searches, analyzes and visualizations.''}

\begin{figure}[!htbp]
  \centering
  \includegraphics[width=\linewidth]{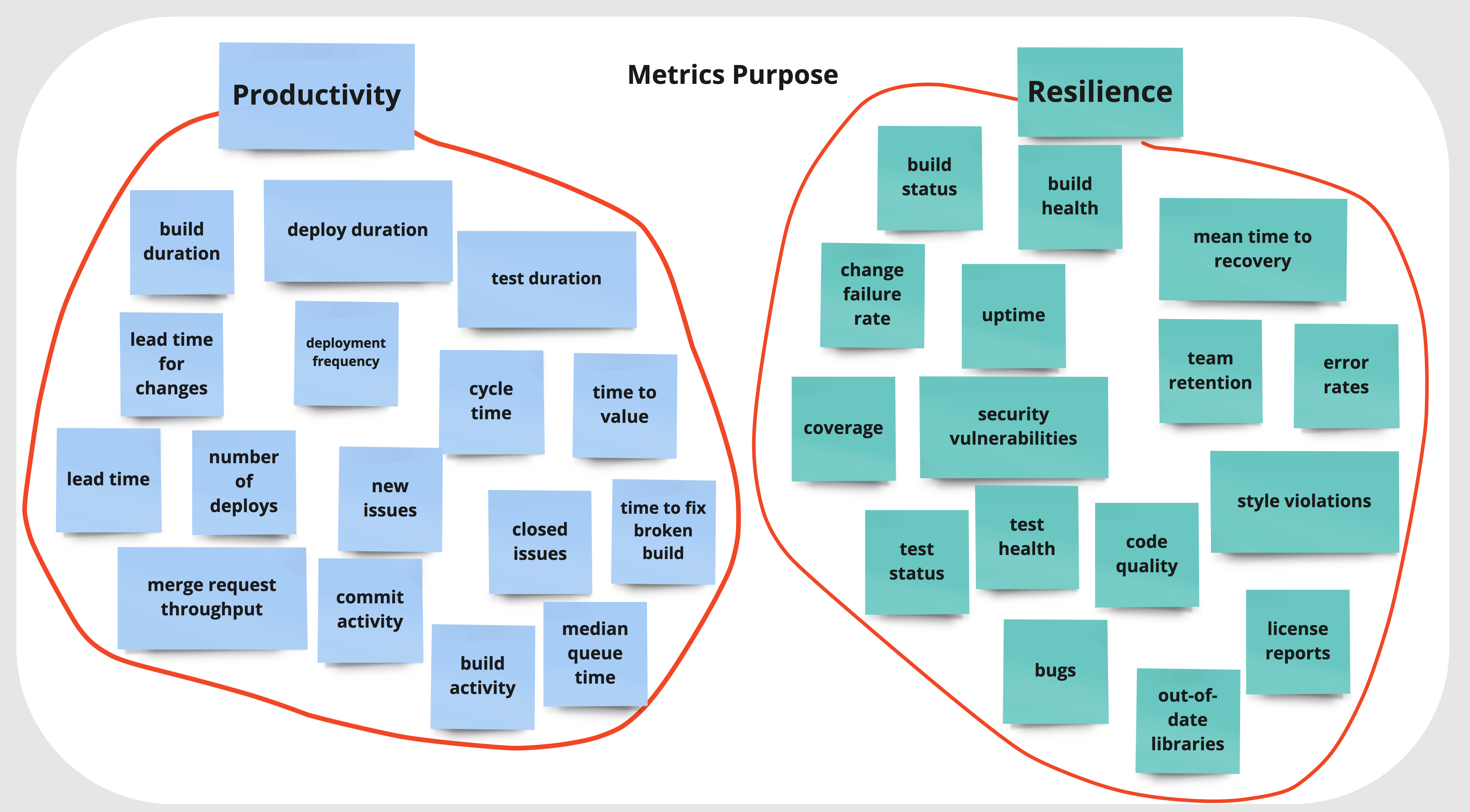}
  \caption{DevOps Metrics purpose}
  \label{fig:practices_purpose}
\end{figure}

In conclusion, our findings indicate that \textsc{Gitlab} stands out as the most comprehensive CI service for monitoring CI practices. While \textsc{DataDog} also offers an extensive monitoring suite. Despite their individual strengths, the combination of these tools, still falls short of fully supporting CI practices monitoring. Moreover, they monitor some identical metrics, such as Coverage, DORA metrics, Build Duration and Build Health. The integration of both tools introduces challenges as it unveils redundant monitoring metrics, resulting in duplicated features and increased financial overhead for companies. Moreover, we realize that we can divide the most of metrics $\frac{32}{33}$ in two groups regarding their purpose: Productivity (Speed) and  Resilience (Quality). Figure  \ref{fig:practices_purpose} outlines this division.

In Table \ref{tab:practices}, we establish a relationship between the DevOps metrics documented in the CI services' official documentation and the seven CI practices defined in the literature by Duvall et al \cite{Duvall2007}. Most of the DevOps metrics can be linked to one or more CI practices defined by Duvall et al \cite{Duvall2007}. For instance, the metrics ``deploy frequency'' and ``cycle time'' are associated with the CI practice defined by Duvall et al as ``Don't commit broken code''. Don't commit broken code helps to increase the ``deploy frequency'' and ``cycle time''. Another example is when you 'Write automated developer tests', which contributes to increasing the ``uptime'' of the application and reduce the ``change failure rate''.

In this way, the current CI services meet the requirements of the CI practices. However, it has been observed that these CI services tend to prioritize higher-level aspects that are more user-focused. For instance, instead of monitoring the frequency of developers' commits, the tools primarily focus on monitoring the frequency of deployments.

\begin{table*}
\caption{Relation between Duvall CI practices \cite{Duvall2007} and DevOps metrics}
\label{tab:practices}
\begin{tabular}{ll}
\hline
\textbf{Duvall CI practice}            & \textbf{DevOps metrics}                                                                                                                                                                                    \\ \hline
Commit code frequently              & \begin{tabular}[c]{@{}l@{}}time to deploy an application, cycle time, time to value, \\ infrastructure costs, deployment frequency, \\ lead time for changes, build activity, number of deploys, \\ closed issues, merge request throughput, lead time \end{tabular}       \\ \hline

Don't commit broken code            & \begin{tabular}[c]{@{}l@{}}time to deploy an application, cycle time, time to value, \\ infrastructure costs, deployment frequency, \\ lead time for changes, build activity, number of deploys, \\ closed issues, merge request throughput, lead time, build health, \\ time to fix broken build, build status, team retention, error rates \end{tabular}    \\ \hline

Fix broken builds immediately       & \begin{tabular}[c]{@{}l@{}}time to deploy an application, cycle time, time to value, \\ infrastructure costs, deployment frequency, \\ lead time for changes, build activity, number of deploys, \\ closed issues, merge request throughput, lead time, build health, \\ time to fix broken build, build status, team retention, error rates \end{tabular}       \\ \hline

Write automated developer tests     & \begin{tabular}[c]{@{}l@{}} time to test an application, time to deploy an application, \\ build health, build duration, lead time for changes, \\  deployment frequency, mean time to recovery, change failure rate, \\ build status, cycle time,  time to value, uptime, error rates, \\ infrastructure costs, team retention,  lead time, \\ coverage, new issues, number of deploys, closed issues, \\ merge request throughput, security vulnerabilities,  code quality, \\ style violations, bugs \end{tabular}        \\ \hline

All tests and inspections must pass & \begin{tabular}[c]{@{}l@{}} build health, build duration, lead time for changes, \\  deployment frequency, mean time to recovery, change failure rate, \\ build status, cycle time,  time to value, uptime, error rates, \\ team retention,  lead time, coverage, \\ merge request throughput, security vulnerabilities,  code quality, \\ style violations, bugs, time to fix broken build \end{tabular} \\ \hline

Run private builds                  & \begin{tabular}[c]{@{}l@{}} time to test an application, time to deploy an application, \\ build health, deployment frequency, lead time for changes, \\ mean time to recovery, change failure rate, \\ build status, cycle time, time to value, error rates, \\ infrastructure costs, team retention, lead time, number of deploys, \\ merge request throughput, build activity, time to fix broken build \end{tabular}                   \\ \hline

Avoid getting broken code           & \begin{tabular}[c]{@{}l@{}}time to test an application, time to deploy an application, \\build health, deployment frequency, lead time for changes, \\ mean time to recovery, change failure rate, \\ build status, cycle time, time to value, uptime, error rates, \\ infrastructure costs, team retention, lead time, number of deploys, \\ merge request throughput, build activity, time to fix broken build \end{tabular}   \\ \hline

\end{tabular}
\end{table*}

\begin{figure}[!htbp]
  \centering
  \includegraphics[width=\linewidth]{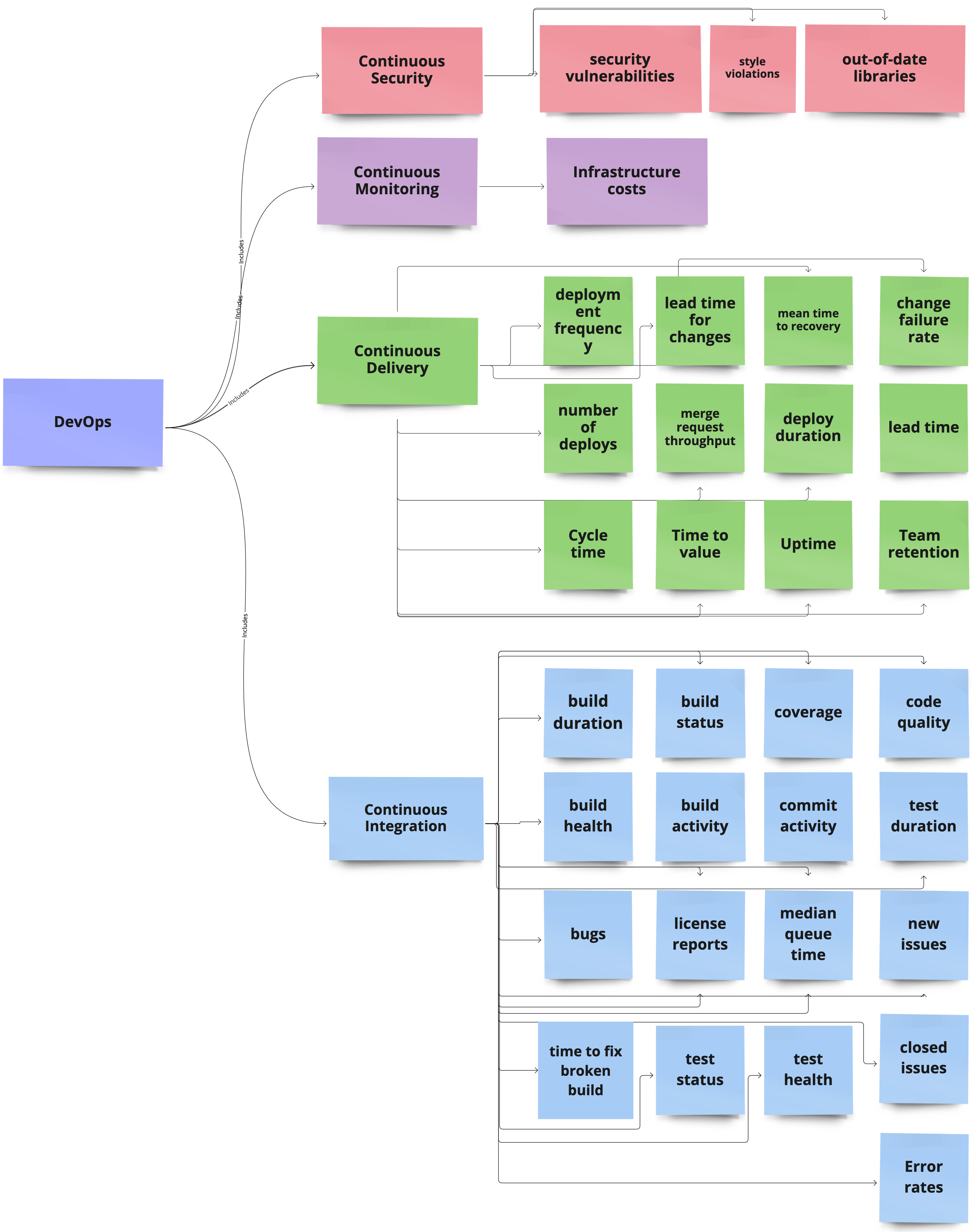}
  \caption{DevOps metrics monitoring overview}
  \label{fig:dev_ops_overview}
\end{figure}

In Figure \ref{fig:dev_ops_overview}, we present an overview divided by DevOps area, showcasing the state-of-the-art monitoring of DevOps metrics compiled from the official documentation of the CI services and third-party monitoring tools.

Lastly, this Research Question reinforces the conclusions of RQ2, indicating that the CI tools still provide preliminary support for monitoring CI practices. Leading developers to use CI tools to monitoring basic build information such as ``build duration'' and ``build status''.

\section{Discussion}  \label{Discussion}

The main goal of our work was to measure developers' perception regarding the monitoring CI practices in their open-source projects. To achieve this, we calculate CI metrics associated with these practices and look for evidence of monitoring these metrics in PR comments, as well as in the documentation of CI services and third-party tools. Additionally, we conducted a survey with developers to explore the evolution of seven CI metrics in their projects. Our data shows that most developers only consider the automatic execution of the CI build and the automatic execution of tests to define the CI maturity level of the project. They found the seven CI metrics presented in our survey important, and they showed interest in monitoring and improving their values. However, as CI services and third-party tools demonstrated only preliminary support to monitor some basic metrics, like build status and build duration, there is a lack of a tool that encompasses all CI metrics. Furthermore, the use of CI services together with third-party tools may generate unnecessary duplicate monitoring of some CI metrics.

We highlight that there is a lack of standardization in the definition and naming of CI (which can also be extended to DevOps) metrics, which makes monitoring CI metrics more complex. In results of RQ3, we provide a catalog of 32 metrics grouped by: (i) CI tools (Figure \ref{fig:themes_ci_servers} and Figure \ref{fig:themes_third_party}), (ii) purpose (Figure \ref{fig:practices_purpose}) and (iii) DevOps area (Figure~\ref{fig:dev_ops_overview}). This catalog can be used by future efforts to standardize these metrics.

We also conducted an assessment of CI services and third-party tools concerning monitoring CI metrics. The data generated by our study can aid in deciding which CI service or third-party tools adopt to monitor these metrics as well as highlights which aspects related to monitoring CI metrics can be improved in future versions of these tools.

Contextual factors such as project size, team structure, and code base complexity undoubtedly play a significant role in influencing the adoption and implementation of Continuous Integration (CI) practices. However, our study deliberately focused on understanding developers' perceptions of CI monitoring practices, regardless of these contextual variables. This approach allowed us to address a critical yet unexplored gap: the disparity between CI adoption and its proper monitoring. As we have shown, previous studies have highlighted that, even when CI is widely adopted, many projects fail to adequately monitor its use, potentially undermining its intended benefits.

Our work makes unique contributions to the literature on CI, complementing and extending previous studies. First, it offers a broader perspective on CI metrics monitoring and its role in ensuring adherence to CI practices. Using two distinct methods across three research questions to triangulate and reinforce our findings, we gather insights from the perspective of project developers and on a larger number of projects than in previous studies. Specifically, our study highlights two key points: (i) The lack of tools designed for the systematic monitoring and reporting of CI metrics. (ii) Identifying developers' interest in monitoring CI practices.

The results of our work highlight a clear lack of tools dedicated to monitoring CI metric to ensure that all CI practices are consistently followed by development teams. Our findings also suggest that developers are interested in monitoring these metrics. Consequently, tools that track CI metrics and provide continuous reports to developers could help identify problems and motivate teams to improve their CI process, ultimately improving project quality. Furthermore, our approach can be applied to closed-source or enterprise projects, as long as a monitoring tool can be developed to collect CI metrics in private repositories and adopted by organizations. The monitoring tool can be flexible and configurable so that the team can choose which CI metrics would be collected and the ideal values of these metrics for a specific project. This is the next step in our research.

Next we discuss the implications of our work for researchers and practitioners.

\textbf{Implications for researchers: }
Our study showed, especially in RQ2, that developers do not monitor CI practices in a systematic manner and do not use clear criteria to define the maturity level of CI in their projects. We believe there is a wide field to better collect and evaluate the impact of these practices. In addition to defining an approach to better measure the level of CI maturity in projects or analyzing other aspects of CI practice monitoring not covered by our study, future research could explore how contextual factors, such as project size, team size, and software complexity, influence the monitoring of CI practices. Furthermore, this study can be extended to include other DevOps practices.

\textbf{Implications for practitioners: } 
Providing clearer strategies, including assigning CI engineers or dedicated monitoring DevOps teams, can help streamline the process and ensure monitoring remains a core part of the development workflow. DevOps teams should incorporate systematic CI monitoring practices into their workflows to ensure continuous feedback on the health of their CI processes. The results of RQ2 indicate that developers consider the seven CI metrics shown in the survey important and would like to monitor them. At the same time, CI tools provide only preliminary support for monitoring CI metrics. Our results suggest that there is a need for CI services to implement or improve the monitoring of CI practices, including increased support for CI metrics monitoring. Effective monitoring tools should be capable of generating automated alerts that notify the team when CI metrics fall below acceptable thresholds, encouraging proactive maintenance of CI health. Regular reports on CI health can further motivate developers by providing a clear view of the project's state and areas for improvement. Additionally, gamification strategies, such as Competition, Rewards, Feedback Loops for maintaining  CI health, can further stimulate developer engagement and encourage consistent best practices. Similarly, standardization of these metrics is necessary. In this sense, initiatives like DORA are very helpful to standardize the definitions. These actions would facilitate and encourage the monitoring of CI practices by project teams.

\section{Threats to Validity} \label{Threats}

In this section, we discuss the threats to the validity of our study.

\textbf{Construct Validity Threats:} The lack of standardization in the definition of CI metrics across different CI tools can lead to errors in interpreting these metrics. In RQ2, survey respondents may respond in a socially desirable way or according to what they think the researcher wants to hear, rather than reflecting their true opinions or feelings. To mitigate this threat, we took the following measures: (i) At the beginning of the survey, we explicitly informed participants that all responses would remain anonymous and would be used solely for academic research purposes, (ii) we carefully formulated survey questions to be neutral, avoiding leading language or phrasing that might suggest a ``preferred'' answer and (iii) we emphasized that participation in the survey was entirely voluntary, and respondents could skip questions or withdraw at any time without any consequences.

\textbf{Internal Validity Threats:} To extract CI information from \textsc{GHActions}, it was necessary to select workflows related to CI. \textsc{GHActions} does not have this information, and as \textsc{GHActions} is a relatively new CI service ( August 2019, \cite{GolzadehCIServer2022}), we did not find a standard way to to extract this type of association in a satisfactory way. Thus, we developed our own heuristic to identify \textsc{GHActions} CI workflows using findings from previous work \cite{DecanGitHubAction2022}. This heuristic may not have covered all cases of CI workflow identification.

A potential limitation of our study is the insufficiency of data for analyzing how contextual factors influence the monitoring of CI practices. While this was not the primary focus of our study, the lack of this data limits the depth of our analysis regarding the broader contextual factors affecting CI monitoring. Future research may aim to collect project-level contextual data to provide a more comprehensive understanding of how these factors influence CI monitoring practices.

Future research should also explore the roles and motivations behind the need for CI practices monitoring within project teams. Understanding who is responsible for tracking CI adoption, enforcing best practices, and interpreting CI metrics can provide valuable insights into how organizations maintain and optimize their CI workflows. This distinction was not explicitly addressed in our study.

For RQ2, we collected data regarding productivity (speed) and quality (resilience) of the development process, using implementation strategies reported by previous works \cite{VasilescuQualityProdOutcomes2015}. To measure productivity, we considered the number of closed PRs, and to measure quality, we considered the number of bug-related issues. However, these measurements represent only a specific perspective of productivity and quality, and other existing perspectives (e.g., effort) should be explored in future work.

We acknowledge the potential for bias in the two Document Analysis performed for RQs 1, and 3, as they relied on the authors' subjective interpretations of PR comments and CI service documentation. We mitigate this threat through the peer review of the codes and themes extracted using Document Analysis. 

\textbf{External Validity Threats: } Despite GitHub and \textsc{GHActions} are widely used in the context of open-source projects, they do not represent the whole of software development scenarios. We acknowledge that our results are restricted to the context of the analyzed projects. Some of our findings, such as the predominance of the \textsc{CodeCov} tool for Coverage monitoring, may not be valid for other contexts.

Although we followed several recommendations \cite{ParticipationSurveys2013} to increase the response rate, we received 28 survey responses, resulting in a response rate of 5.6\%. However, we consider these responses valuable as they provide insights into the opinions and practices of core developers with extensive knowledge of the CI practice in the analyzed projects. The response rate achieved can be attributed to our focus on selecting qualified developers with expertise in the CI field. To mitigate the impact of the response rate, we conducted two additional analyses and triangulated the results with those from the survey. To generalize the results for projects of different natures, further replication studies with a larger number of developers are needed.

\section{Conclusion} \label{Conclusion} 

In this paper, we presented a qualitative study on developers' perceptions of monitoring CI practices. To achieve this, we conducted a Document Analysis to discover the current needs for monitoring CI practices expressed by developers during the development process, using pull request comments. We then collected the evolution of seven CI metrics associated with CI practices over a period of 2 years from 121 open-source projects and conducted a survey with 28 developers from these projects. The goal was to understand the importance of the collected practices to the developers and to gain insights into how they measure the adoption of CI in their projects and their needs related to CI monitoring. Finally, we performed another Document Analysis to assess the current level of support provided by CI services for monitoring CI practices.

Our main results were as follows: 1)  Coverage is the most monitored DevOps metrics, and \textsc{CodeCov} is the most used tool by developers during the development process. 2) There is no standard set of practices or data that developers use to define the maturity level of their CI projects. 3) Most developers found the seven CI metrics presented in our survey important and expressed interest in monitoring them if provided the opportunity. 4) CI tools have preliminary support to monitor CI metrics, and GitLab is one platform that started to monitor some metrics, however with focus on DevOps metrics like DORA. 5) We also found a need to monitor ``Build health'' and ``Time to fix a broken build'' metrics.

\section*{Declarations}

\subsection*{Ethical Approval}

Not Applicable

\subsection*{Informed Consent}

Informed consent was obtained from all participants prior to completing the survey.

\subsection*{Author Contributions}

The first author designed the study, collected data, sent the survey to the participants, and performed the analysis of the results. The second author reviewed the document analysis for RQ1 and RQ3 as well as the manuscript. The third author reviewed the manuscript. The fourth author re-evaluated disagreements between the first and second authors regarding the document analysis of RQ3 and also reviewed the manuscript.

\subsection*{Data Availability Statement}

The authors declare that they store all data searched and collected in an external public repository available at \textbf{https://doi.org/10.5281/zenodo.14569025}.

%
\subsection*{Competing Interests}

The authors declare that they have no conflicts of interest or competing interests.

\subsection*{Funding}

This work received no funding.


\bibliographystyle{spmpsci}      
\bibliography{main}


\end{document}